\documentclass[sigconf]{acmart}

\copyrightyear{2026}
\acmYear{2026}
\setcopyright{cc}
\setcctype{by}
\acmConference[ICSE '26]{2026 IEEE/ACM 48th International Conference on Software Engineering}{April 12--18, 2026}{Rio de Janeiro, Brazil}
\acmBooktitle{2026 IEEE/ACM 48th International Conference on Software Engineering (ICSE '26), April 12--18, 2026, Rio de Janeiro, Brazil}
\acmPrice{}
\acmDOI{10.1145/3744916.3773166}
\acmISBN{979-8-4007-2025-3/2026/04}

\usepackage{amsfonts} %
\usepackage{algorithmic} %
\usepackage{textcomp} %
\usepackage{fancybox} %
\usepackage{amssymb} %
\usepackage{stfloats} %
\usepackage{calc} %
\usepackage{lscape} %
\usepackage{verbatim} %
\usepackage{makecell} %
\usepackage{bbding} %
\usepackage[commandnameprefix=always,defaultcolor=red]{changes} %
\usepackage[labelformat=simple]{subcaption} %
\usepackage{bm} %
\usepackage{framed}
\usepackage{mathtools} %
\usepackage{pifont} %
\usepackage{soul} %
\usepackage{enumitem} %
\usepackage{siunitx} %
\usepackage{multirow} %
\usepackage{listings} %
\usepackage{amsthm} %
\usepackage{url} %
\usepackage{colortbl} %
\usepackage{color} %

\hypersetup{
 colorlinks=true,
 linkcolor=purple,
 citecolor=violet,
 filecolor=magenta,      
 urlcolor=cyan,
}

\newcommand{\PP}[1]{
\vspace{2px}
\noindent{\bf \IfEndWith{#1}{.}{#1}{#1.}}
}

\newcommand{\PO}[1]{
\vspace{2px}
\noindent{\bf \IfEndWith{#1}{.}{#1}{#1:}}
}

\newcommand{\code}[1]{{\fontfamily{cmtt}\fontseries{b}\fontshape{h}\selectfont\small{#1}}}

\DeclareRobustCommand{\relscale}{\originalrelscale}
\let\originalrelscale\relscale

\definecolor{verylightgray}{rgb}{.95,.95,.95}

\lstdefinelanguage{Solidity}{
    identifierstyle=\color{black},
    sensitive=false,
    comment=[l]{//},
    morecomment=[s]{/*}{*/},
    commentstyle=\color{gray}\footnotesize\ttfamily,
    stringstyle=\color{black}\ttfamily,
    morestring=[b]',
    morestring=[b]"
}

\definecolor{mycustomcolor}{HTML}{175c89}   
\definecolor{ToolNameColor}{HTML}{040e69}
\newcommand{\ToolName}{\textcolor{ToolNameColor}{\textsf{{EchoFuzz}}}}
\definecolor{UrlColor}{HTML}{3300FF}

\setlist[itemize]{leftmargin=*}
\setlist[enumerate]{leftmargin=*}
\newlist{steps}{enumerate}{1}
\setlist[steps, 1]{label = \textbf{RQ\arabic*.}}

\AtBeginDocument{%
  }

\begin{document}
\begin{sloppypar}

\title{EchoFuzz: Empowering Smart Contract Fuzzing with Large Language Models}

% the first author
\author{Juanen Li}
\affiliation{%
  \institution{Tsinghua University \& Beihang University \& Beijing Normal University}
  \city{Beijing}
  \state{}
  \country{China}
}

% second author
\author{Peng Qian}
\affiliation{%
  \institution{Zhejiang Gongshang University \& Goplus Security}
  \city{Hangzhou}
  \state{}
  \country{China}}
\email{}

% third author
\author{Guanyan Li}
\affiliation{%
  \institution{University of Oxford}
  \city{Oxford}
  \country{United Kingdom}
}

% fourth author
\author{Rui Wang}
\affiliation{%
 \institution{Beijing Normal University}
 \city{Zhuhai}
 \state{}
 \country{China}
}

% fifth author
\author{Peixin Wang}
\affiliation{%
  \institution{East China Normal University}
  \city{Shanghai}
  \state{}
  \country{China}
}

% sixth author
\author{Zhiqing Tang}
\affiliation{%
  \institution{Beijing Normal University}
  \city{Zhuhai}
  \state{}
  \country{China}}
\email{}

% seventh author
\author{Fuchen Ma}
\affiliation{%
  \institution{Tsinghua University}
  \city{Beijing}
  \country{China}}
\email{}

% eighth author
\author{Yuanliang Chen}
\affiliation{%
  \institution{Tsinghua University}
  \city{Beijing}
  \country{China}}
\email{}

\authornote{\textit{Yuanliang Chen is the corresponding author.}}

% ninth author
\author{Lun Zhang}
\affiliation{%
  \institution{GoPlus Security}
  \city{Hangzhou}
  \country{China}
}
\email{}

\begin{abstract}
Smart contracts, serving as the cornerstone of decentralized applications, autonomously manage trillion-dollar digital assets, making them attractive targets for attacks. Fuzzing has emerged as a promising technique for detecting vulnerabilities in smart contracts, yet existing methods face two main challenges. (1) The logical gap in state transitions and combinatorial redundancy hinders effective tradeoffs between bug detection efficiency and state space exploration cost, leading to critical execution paths to be overlooked. (2) Rule-based sequence mutation strategies suffer from path redundancy and inadequate guidance from contract logic, resulting in performance bottlenecks that stall the exploration of in-depth vulnerability-oriented paths.

To tackle these challenges, we propose \ToolName{}, an LLM-guided fuzzing framework introducing Vulnerable Function Call Sequences (VFCS) - minimal, behavior-preserving execution paths that expose bugs through key state transitions. \ToolName{} consists of two key procedures. First, we develop a chain-guided LLM approach, that combines static analysis with logical understanding to generate contract-specific VFCS candidates that eliminate combinatorial redundancy. Second, we adopt an iterative fuzzing strategy that uses LLMs with real-time feedback to adaptively steer fuzzer toward uncovered branches. Experiments show \ToolName{} outperforms state-of-the-art methods, achieving 29\% higher branch coverage and detecting 62\% more vulnerabilities. It also found 37 previously unknown vulnerabilities in real contracts, showing strong practicality.

\end{abstract}


\maketitle

\section{Introduction}
Smart contracts are blockchain-based programs that autonomously execute predefined terms, serving as the cornerstone of decentralized applications across various domains~\cite{kosba2016hawk,chen2023tyr,ren2021making}. 
Pioneering platforms like Ethereum host millions of contracts~\cite{wood2014ethereum,ma2023pied,ma2021security}, revolutionizing domains like digital finance~\cite{antonopoulos2018mastering,dai2021trustzone}, supply chain~\cite{pawar2021secure,bains2022regulating}, and crowdfunding~\cite{jacynycz2016betfunding,zhang2024nyx}.
They now manage trillion-dollar assets, but vulnerabilities have caused over \$10 billion~\cite{pigni2018targeting,hasan2018proof,zheng2020overview} in losses, underscoring the critical need for robust security~\cite{newman2009computer}.

To ensure the security of smart contracts, a variety of vulnerability detection techniques are employed~\cite{amankwah2017evaluation,chakraborty2021deep}. Among these, fuzzing~\cite{li2018fuzzing} is considered one of the most powerful and encouraging methods for automated vulnerability detection~\cite{russell2018automated,li2016vulpecker}. 
Fuzzing detects vulnerabilities in smart contracts by repeatedly generating and executing a wide range of random or targeted inputs to test the contract's functions. During this process, the fuzzer is able to explore various execution paths within the contract to identify abnormal or unexpected behaviors~\cite{bonett2018discovering}.

However, due to its transaction-centric nature, fuzzing smart contracts presents unique challenges.
Heuristically, the essence of smart contract fuzzing is to explore the possible execution sequences, i.e., [$f_1(\mathtt{a}_1) \rightarrow f_2(\mathtt{a}_2) \rightarrow \dots \rightarrow f_n(\mathtt{a}_n)$], where $f_i()$ denotes a function call in the contract and $\mathtt{a}_i$ is its argument. 
The goal is to check if the final state reveals vulnerabilities with some given oracles.
Despite advances in argument generation, creating quality call sequences remains hard due to vulnerability interdependencies on call orders.
Poorly designed sequences may fail to expose complex vulnerabilities, as rigid patterns restrict access to essential execution paths.
Simple sequences ignore contract logic, while full transaction call record simulation is computationally expensive, hindering the exploration of in-depth and hidden vulnerabilities. Thus, generating effective sequences faces two key challenges.

\phantomsection
\label{sec:challenge1}
\PP{Challenge 1: Balancing Vulnerability Detection Probability and Exploration Space Cost}
State space explosion arises from software engineering limitations: logical gaps in state transitions and combinatorial redundancy.
Traditional methods track data dependencies but miss key functional relationships like state-offsetting interactions.
This creates a trade-off between exploration cost and detection probability.
Oversimplified state spaces miss critical paths, while over-extended ones lead to high computational cost without significant gains in vulnerability detection.

\phantomsection
\label{sec:challenge2}
\PP{Challenge 2: Evolving State Space Exploration Beyond Fuzzing Bottlenecks.} 
Fuzzing often plateaus, with fewer new vulnerabilities found over time.
Conventional methods either continue within the same state space or use rule-based mutations, which often produce redundant sequences.
These approaches revisit previously explored areas instead of probing new, underexplored regions. 
Without logic-driven adaptation, they lack targeted exploration, reducing efficiency and new vulnerability discovery.

\begin{figure*} [t]
    \centering
    \includegraphics[width=1\linewidth]{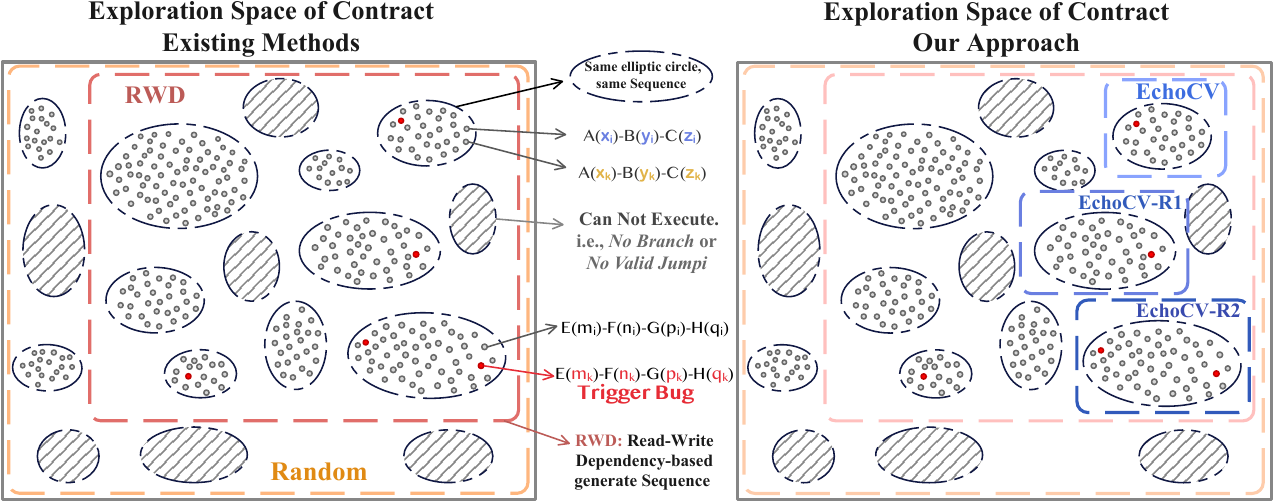}
    \Description{}
    \vspace{-16pt}
    \caption{\small Each ellipse represents a specific sequence, and the points within the ellipses denote test cases with parameters that into fuzzing; \textcolor{gray}{gray points} indicate test cases that execute normally, while \textcolor{red}{red points} represent test cases that reveal vulnerabilities. The \textcolor{gray}{gray-shaded} areas within the ellipses signify sequences that are non-executable, labeled as \emph{No Branch} or \emph{No Valid Jumpi}.}
    \label{fig:intro-ab}
    \vspace{-10pt}
\end{figure*}

Existing methods struggle to address these challenges effectively.
As in Figure~\ref{fig:intro-ab}, random generation techniques (orange box) produce arbitrary outputs rarely finding meaningful patterns in complex contracts.
Read-write dependency-based (RWD) methods (maroon box) analyze sequence dependencies via contract read/write operations.
While RWD methods improve on random generation by filtering invalid sequences, their rule-based approaches cause overgeneralization and imprecision with contract-specific logic.
RWD methods, such as constraint-based symbolic execution~\cite{torres2021confuzzius}, reduce the exploration space by enforcing constraints, but they are computationally expensive and struggle with non-deterministic behavior.
Similarly, other RWD approaches, including dynamic taint~\cite{ji2023seqfuzz} and evolutionary algorithms~\cite{li2024evofuzzer}, overproduce scenarios but miss nuanced vulnerabilities. 
These methods often overlook in-depth bugs, converge poorly, and lack adaptability. 
Recent attempts using reinforcement learning~\cite{su2022effectively} also struggles with defining robust reward functions and generalizing across diverse contract types.

\PP{Our Approach.}
To address these challenges, we introduce \ToolName{}, 
a fuzzing framework that leverages Large Language Models (LLMs)~\cite{zhao2023survey} for logical reasoning.
Unlike traditional methods, 
it uses LLM-guided strategies to deeply explore smart contract state spaces.
Central to our approach is the introduction of Vulnerable Function Call Sequences (VFCS), a novel abstraction that identifies minimal, logically coherent paths to directly trigger vulnerabilities, overcoming problems such as the logical gap and combinatorial redundancy. VFCS is defined by a series of dependent function calls that target areas most likely to contain vulnerabilities while minimizing unnecessary exploration. Inspired by expert analysis, we use LLMs to interpret code and predict vulnerable paths via state transitions. The iterative nature of our approach, guided by real-time feedback, ensures efficient vulnerability detection. \ToolName{} integrates two critical components that guide the fuzzing process through contract logic analysis and iterative refinement to efficiently and effectively identify vulnerabilities.

\PP{Procedure 1: Chain-Guided Candidate VFCS Generation}
For \emph{Challenge 1}, we generate candidate VFCSs (cf. Section~\ref{sec:vfcs-definition} for details) via an LLM-driven chain-guided process.
Using code comprehension, static analysis, and few-shot prompting, the LLM targets high-risk paths and key state transitions to reveal vulnerabilities.
This context-aware method narrows the exploration space, ensuring critical VFCSs are found.
Experiments show this method generates targeted sequences, improving exploration efficiency and vulnerability detection.
As shown in Figure~\ref{fig:intro-ab}, {\textcolor{ToolNameColor}{\textsf{{\underline{Echo}Fuzz}}}} generates \underline{c}andidate \underline{V}FCSs (denoted as \textbf{EchoCV}, the blue dashed box) through LLM-guided analysis, capturing areas with the highest likelihood of vulnerabilities.

\PP{Procedure 2: Feedback-Driven LLM-Guided Iterative Fuzzing}
For \emph{Challenge 2}, we use an iterative fuzzing strategy with real-time feedback and LLM-based code comprehension.
It analyzes runtime feedback to find unexplored or poorly covered paths.
Based on this feedback, the LLM analyzes the logic of branch patterns and generates focused sequences to drive the fuzzer towards the uncovered areas, preventing stagnation during exploration. 
As in Figure~\ref{fig:intro-ab}, this iterative approach generates new candidate sequences, EchoCV-N1 and EchoCV-N2, targeting previously uncovered branches, enhancing coverage and vulnerability detection.

We conduct experiments to evaluate the effectiveness of our proposed framework \ToolName{} across three datasets, including large-scale real-world contract blockchain projects. 
Empirical results strongly support our claim that \ToolName{} mitigates key challenges in smart contract fuzzing.
Specifically, \ToolName{} achieves an average improvement in 29.19\% in branch coverage and detects 62.77\% more vulnerabilities compared to the current state-of-the-art. Encouragingly, \ToolName{} uncovers 37 previously unknown vulnerabilities in various contract projects, showcasing its strong performance and practical value in enhancing security.

\PP{Main Contributions.}
To summarize, the key contributions of this work are as follows:
\begin{itemize}[topsep=1.5pt, itemsep=0pt, left=0.2em] 
    \item We present \ToolName{}, the first fuzzing framework to use chain-guided LLMs for smart contract logic, focusing on vulnerable function call sequences to pinpoint the paths that trigger vulnerabilities, overcoming the limitations of traditional methods.
    \item \ToolName{} combines contextual and static analysis to enhance VFCS generation and uses a real-time feedback-driven, iterative fuzzing process to dynamically refine strategies, expanding coverage and boosting vulnerability detection. 
    \item Experimental results show that \ToolName{} outperforms current state-of-the-art fuzzers in code coverage and vulnerability detection. In the spirit of open science
    \footnote{\textcolor{UrlColor}{\url{https://github.com/iceray00/EchoFuzz}}}
    , we release our framework for further research and development. 
\end{itemize}

\section{Background}

\PP{Ethereum Smart Contract.}  
Smart contracts, conceptualized by Nick Szabo in 1994~\cite{cutts2019smart}, are self-executing programs on blockchains like Ethereum that enforce predefined terms without intermediaries.
These attributes enable their application in decentralized finance (DeFi)~\cite{zetzsche2020decentralized,werner2022sok}, voting systems~\cite{nurmi2012comparing}, and non-fungible tokens (NFTs)~\cite{chohan2021non,sestino2022non,valeonti2021crypto}. 
However, the substantial digital assets they hold attract sophisticated attacks, causing billion-dollar losses, which makes robust vulnerability detection crucial. 

\PP{Smart Contract Fuzzing.}
Unlike traditional fuzzing that targets crashes, smart contract fuzzing must address state-dependent execution, as contract state relies on prior transaction sequences~\cite{hart1988incomplete,hart1986theory}.
Thus, its core task is generating and mutating transaction sequences to explore these state dependencies.
Furthermore, vulnerabilities in smart contracts often stem from complex logic errors rather than runtime crashes. In this context, fuzzers typically integrate predefined oracles within the execution engine 
to detect specific flaw patterns like reentrancy or block dependency~\cite{wloka2009refactoring,idelberger2016evaluation,huang2019smart}.

\PP{LLM for Fuzzing in Smart Contracts.}
Large Language Models (LLMs)~\cite{huang2022large,zhao2023survey} are advanced AI systems with strong logical reasoning and text generation capabilities~\cite{yang2019exploring}.
These capabilities enable their use in fuzzing frameworks~\cite{shou2024llm4fuzz,deng2023large}, where they enhance vulnerability detection by generating realistic inputs and contextually relevant transaction sequences that reflect genuine user interactions~\cite{sravanthillms}.
This overcomes limitations of traditional methods, which often rely on random or rule-based generation, lack contract-specific logic, and miss nuanced vulnerabilities~\cite{shuaib2018using}.
Leveraging LLMs thus enables more sophisticated fuzzing techniques for accurate vulnerability detection, strengthening smart contract security.
 
\section{Motivating Example}

\label{sec:motivating-example}

\begin{figure} [t]
    \centering
    \includegraphics[width=0.98\linewidth]{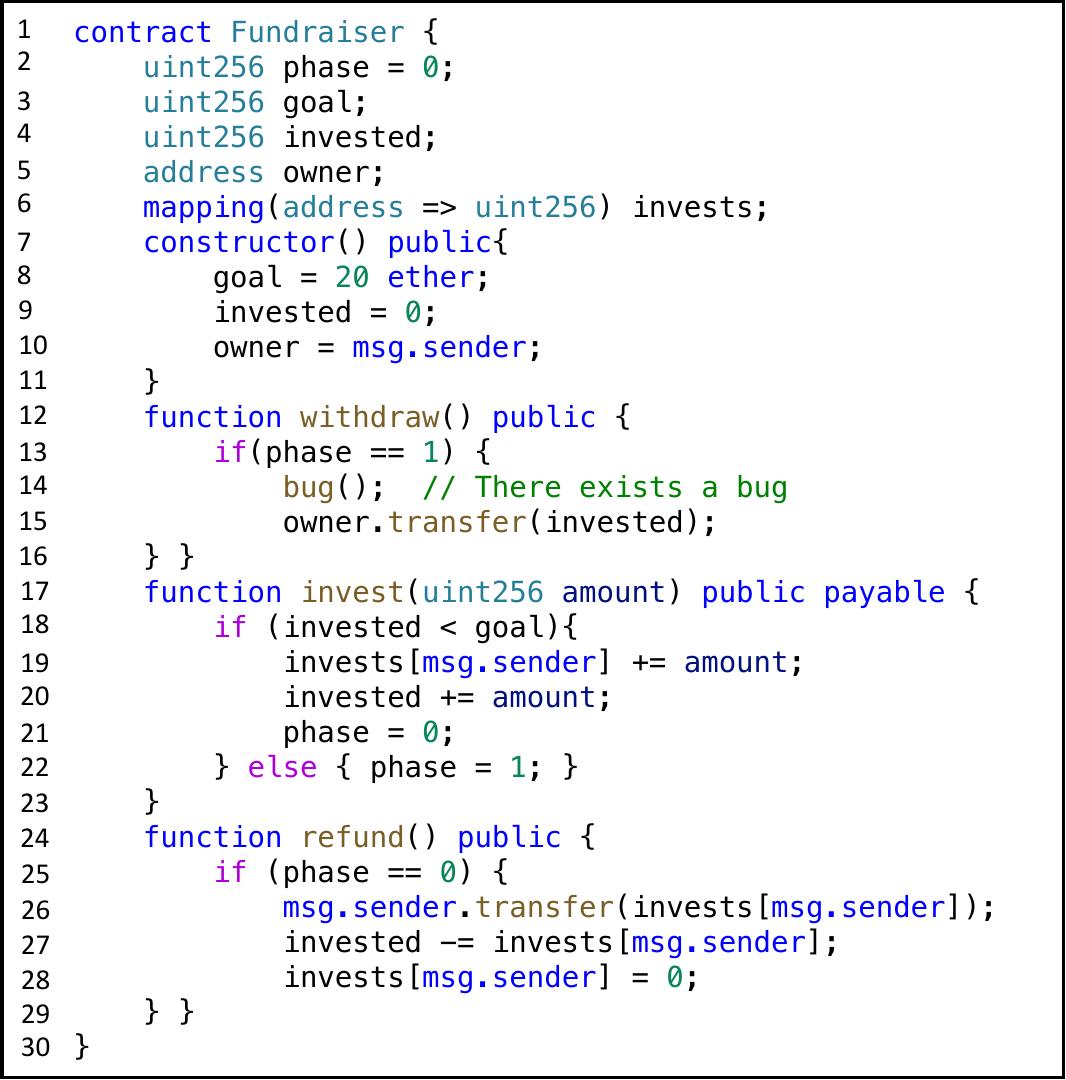}
    \vspace{-8pt}
    \caption{\small Fundraiser, a Loan System Solidity Code~\cite{qian2024mufuzz}}
    \label{fig:crowdsale}
    \vspace{-12pt}
\end{figure}

To further illustrate our idea, let us present a more detailed example. Consider the motivating example shown in Figure~\ref{fig:crowdsale}, a simplified version of the \code{Fundraiser}, which is a variant of~\cite{qian2024mufuzz} -- abstracted from a real-world contract to capture its core functions and bug.

The \code{Fundraiser} contract consists of a \code{constructor} and three key functions: \code{withdraw}, \code{invest}, and \code{refund}. The \code{constructor} is executed only once when the contract is deployed. It sets the fundraiser goal to 20 ether (i.e., \code{goal} = 20 ether on line 8), initializes the invested amount to 0 (i.e., \code{invested} = 0 on line 9), and designates the contract creator as the owner (i.e., \code{owner} = \code{msg.sender} on line 10). Users can participate in the \code{Fundraiser} by calling \code{invest} while the contract is active (i.e., \code{phase} = 0). During this period, users can request a refund by invoking \code{refund} while the \code{Fundraiser} is ongoing (\code{phase} = 0). The \code{Fundraiser} concludes when the investment reaches or exceeds the goal (i.e., \code{invested} $\geq$ 20), at this point the contract owner can withdraw all funds by invoking the \code{withdraw} function. In this context, the function \code{withdraw} contains a bug at line 14 that can only be detected by the fuzzer if the branch condition satisfies \code{phase} = 1. 
Achieving this goal is by no means trivial, as it requires the fuzzer to generate executable transaction sequences while accounting for strict dependencies and the impact of other transactions, which is often quite challenging.

\begin{figure}[t]
    \centering
    \includegraphics[width=\linewidth]{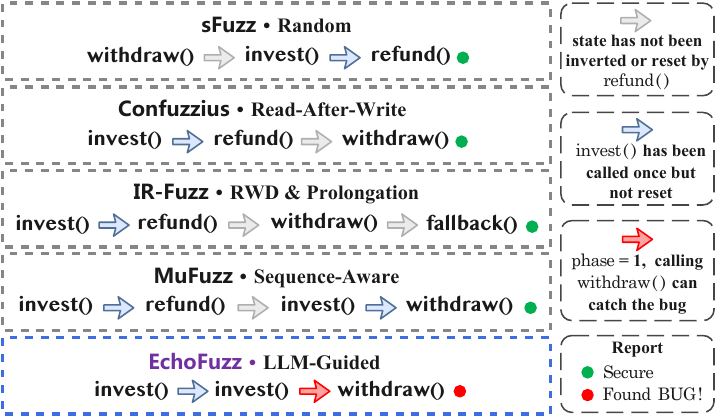}
    \vspace{-18pt}
    \captionof{figure}{\small Different fuzzers generate sequences}
    \label{fig:moe-diff}
    \vspace{-10pt}
\end{figure}

For instance, suppose that the fuzzer generates a sequence [\code{withdraw} $\to$ \code{invest}($\ast$) $\to$ \code{refund}], where $\ast$ can be any value, meaning all situations in the function \code{invest} can bring. 
However, due to a read-dependency of the state variable \code{phase} on \code{invest}, mutating inputs within this sequence cannot trigger the bug.
To catch the bug, the fuzzer may need to generate a sequence such as [\code{invest}($\ast$) $\to$ \code{refund} $\to$ \code{withdraw}]. However, this sequence also fails to catch the bug, as calling \code{invest} once cannot enter the else-branch at line 22 to set \code{phase} = 1. To reach the else-branch, \code{invest} needs to be executed at least twice in the execution sequence, such as [\code{invest}($\ast$) $\to$ \code{refund} $\to$ \code{invest}($\ast$) $\to$ \code{withdraw}]. However, this sequence overlooks the impact from the function \code{refund} -- once \code{refund} is called after \code{invest}, the state variable \code{invests[msg.sender]} resets to 0, making \code{invested} be reduced, as a result offsetting the effect of the previous \code{invest} call. Hence, sequences like [\code{invest}($\ast$) $\to$ \code{refund} $\to$ \code{invest}($\ast$) $\to$ \code{withdraw}] or [\code{invest}($\ast$) $\to$ \code{invest}($\ast$) $\to$ \code{refund} $\to$ \code{withdraw}] cannot set \code{phase} = 1, which fails to catch the bug at line 14. Furthermore, to find the critical sequence that catches the bug while reaching a minimized state space, the ideal sequence would be [\code{invest}($\ast$) $\to$ \code{invest}($\ast$) $\to$ \code{withdraw}]. Without calling the \code{refund} function, the first call meets the fundraiser goal, and the second enters the else-branch at line 22. Next, we show the limitations of existing fuzzers for identifying such a specific sequence, which we call the vulnerable function call sequence.

{
\PP{Limitations of Existing Fuzzers} 
We evaluate several state-of-the-art smart contract fuzzers to our example, including sFuzz~\cite{nguyen2020sfuzz}, Confuzzius~\cite{torres2021confuzzius}, IR-Fuzz~\cite{liu2023rethinking}, and MuFuzz~\cite{qian2024mufuzz}. 
None detect the bug at line 14, as they fail to generate sequences that enter the if-branch at line 13.
In contrast, \ToolName{} reasons about the contract logic, entering the critical branch and exposing the bug in seconds. Existing fuzzers struggle to identify critical sequences and reach deep branches. As Figure~\ref{fig:moe-diff} shows, they typically call all functions for completeness and order them via predefined rules.

Specifically, sFuzz uses a random generation strategy, making it difficult to identify meaningful sequences. 
Confuzzius utilizes the read-after-write technique and combines symbolic execution for dynamic data analysis to find potentially relatively meaningful sequences, but struggles to handle situations where a function must be called multiple times, as the example in Figure~\ref{fig:crowdsale}.
IR-Fuzz employs static data flow analysis to prioritize and introduces a prolongation technique, but still struggles with calling the same function multiple times. MuFuzz focuses on mutating sequences and detects that \code{invested} < \code{goal} is the key to branching, but it fails to account for the impact of \code{refund} on the contract state, missing interaction cancellation, and thus the vulnerability.

These limitations indicate that existing fuzzers tend to generate smaller state spaces to ensure computation feasibility and fuzzing efficiency, thus preventing excessively large state spaces in large contracts, which complicates the analysis. This hinders the ability to reach deeper path branches, preventing the discovery of complex vulnerabilities (corresponding to the aforementioned \hyperref[sec:challenge1]{\emph{Challenge 1}}). Furthermore, the mutation technique used by fuzzers is constrained by predefined rules. While aiming for universality in sequence mutations, they inevitably overlook the unique logical designs of specific contracts. As a result, this limitation obstructs effective navigation toward subsequent meaningful path branches, ultimately creating a frustrating bottleneck within the fuzzing process (corresponding to \hyperref[sec:challenge2]{\emph{Challenge 2}}). In the following sections, we will introduce our innovative methodology to address these challenges.
}

\section{Methodology}
\begin{figure}[t]
    \centering
    \vspace{-0pt}
    \includegraphics[width=\linewidth]{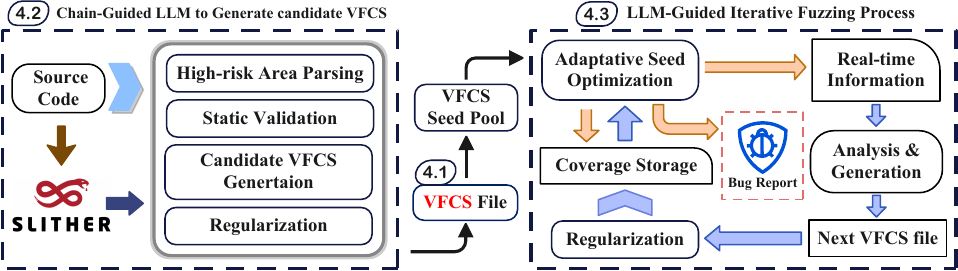}
    \vspace{-17pt}
    \caption{\small A high-level overview of \ToolName{}, including two key procedure: (1) Chain-guided LLM to Generate candidate VFCS and (2) LLM-guided Iterative Fuzzing Process.}
    \label{fig:overview}
    \vspace{-13pt}
\end{figure}

\PP{Overview} The overall architecture of our framework \ToolName{} is outlined in Figure~\ref{fig:overview}. To address the above challenges, our framework shifts the fuzzing strategy from arbitrary function compositions to Vulnerable Function Call Sequences (VFCS). This approach is designed to:
\begin{itemize}
    \item  Balance vulnerability detection probability and exploration cost by generating minimal, logically meaningful sequences.
    \item  Overcome fuzzing stagnation through LLM-guided iterative refinement of sequences.
\end{itemize}
We first formalize VFCS as the core abstraction for vulnerability-centric exploration, then detail how \ToolName{} generates and optimizes these sequences using LLMs. The \texttt{Fundraiser} contract (Figure~\ref{fig:crowdsale}) is used throughout this section to illustrate our method.

\subsection{Vulnerable Function Call Sequence (VFCS)} \label{sec:vfcs-definition}

Existing software engineering-based techniques, such as symbolic execution, static analysis and other rule-based mutation, face critical shortcomings in smart contract fuzzing: (1) Logical Gap in State Transitions. While they can track inter-function data dependencies (e.g., invested is read in withdraw), they fail to recognize functional relationships like \emph{offsetting interactions}. For example, in the \texttt{Fundraiser} contract, \code{refund} resets \code{invested}, neutralizing the cumulative effect of prior \code{invest} calls -- a nuance overlooked by rule-based approaches. (2) Combinatorial Redundancy. Tools like MuFuzz generate redundant sequences (e.g., [\code{invest} $\to$ \code{refund} $\to$ \code{invest} $\to$ \code{withdraw}]) to further increase the probability of discovering vulnerabilities, significantly increasing computational costs compared to minimal paths. These shortcomings exemplify the fundamental tension in \emph{Challenge 1}, i.e. achieve a high detection probability without prohibitive exploration costs.

To address these limitations, we propose \textbf{Vulnerable Function Call Sequence (VFCS)} as a logic-aware paradigm. VFCS tackles these issues by formalizing minimal critical paths that are directly responsible for triggering vulnerabilities. A VFCS is defined as a sequence \( \nu = [f_{\nu_1}, f_{\nu_2}, \ldots, f_{\nu_k}] \) with three essential properties:

\PP{1} \textbf{Vulnerability Trigger:} The final call \( f_{\nu_k} \) executes under a state \( \mathcal{S}_k \) that exposes the vulnerability.

\PP{2} \textbf{Dependency Chain:} Each \( f_{\nu_i} (i < k) \) alters \( \mathcal{S} \) to enable \( f_{\nu_{i+1}} \). Formally:
    \vspace{-5pt}
   \[
   \mathcal{S}_i \xrightarrow{f_{\nu_i}} \mathcal{S}_{i+1} \quad \text{where} \quad \mathcal{S}_{i+1} \models \text{Pre}(f_{\nu_{i+1}})
   \]
\PP{3} \textbf{Minimality:} No proper subsequence of \( \nu \) can trigger the bug.  

The formula \( \mathcal{S}_i \xrightarrow{f_{\nu_i}} \mathcal{S}_{i+1} \) represents the transition of the system state \( \mathcal{S} \) after executing function \( f_{\nu_i} \), where \( \mathcal{S}_{i+1} \) satisfies the preconditions for the next function call. The symbol \( \models \) means ``satisfies'', indicating that the new system state \( \mathcal{S}_{i+1} \) meets the preconditions for the next function call \( f_{\nu_{i+1}} \).

For instance, the sequence [\code{invest} $\to$ \code{invest} $\to$ \code{withdraw}] in motivating example is a VFCS, because: (1) \code{withdraw} executes when \code{phase = 1}, triggering the bug; (2) For dependency chain, first \code{invest}, sets \code{invested = donations}. Second \code{invest}, pushes \code{invested $\ge$ goal}, transitioning \code{phase = 1}; (3) Minimality: Inserting \code{refund} or removing either \code{invest} breaks the chain.

However, directly generating such sequences is impractical, and satisfying all VFCS properties simultaneously is idealized. Without prior knowledge of vulnerability locations and exact trigger conditions, simultaneously ensuring minimality, maintaining dependency chains, and triggering specific vulnerabilities is impractical in practice. Hence, we pursue VFCS candidates step by step that can approximate three properties as closely as possible, thereby reducing the gap between the candidate VFCS and the ideal concept.

\subsection{{\fontsize{10.1}{10}\selectfont Chain-Guided LLM to Generate Candidate VFCS}}
\label{sec:vfcs-generation}

To bridge this gap, we take advantage of the code comprehension capabilities of LLMs to generate candidate VFCS that better approximate these ideal properties.

\textbf{How can we guide LLMs to achieve this goal?} To answer this, we abstract the workflow of expert contract auditing to develop a structured approach for sequence extraction, consisting of three key steps:
\textit{(1) High-risk Area Identification.} Identifying critical functions and variables, taking into account potential state alterations due to function offsetting relationships (e.g., \code{invest} and \code{refund} in \texttt{Fundraiser} which modify shared state variables like \code{phase}). 
\textit{(2) Contextual Validation.} Validating hypotheses using static analysis tools (e.g., control flow graphs from \emph{Slither}) to establish execution constraints. 
\textit{(3) Pattern Matching.} Leveraging known vulnerability patterns to guide scenario construction (e.g., identifying potential bugs in \code{withdraw}).

\begin{figure} [t]
    \centering
    \includegraphics[width=1\linewidth]{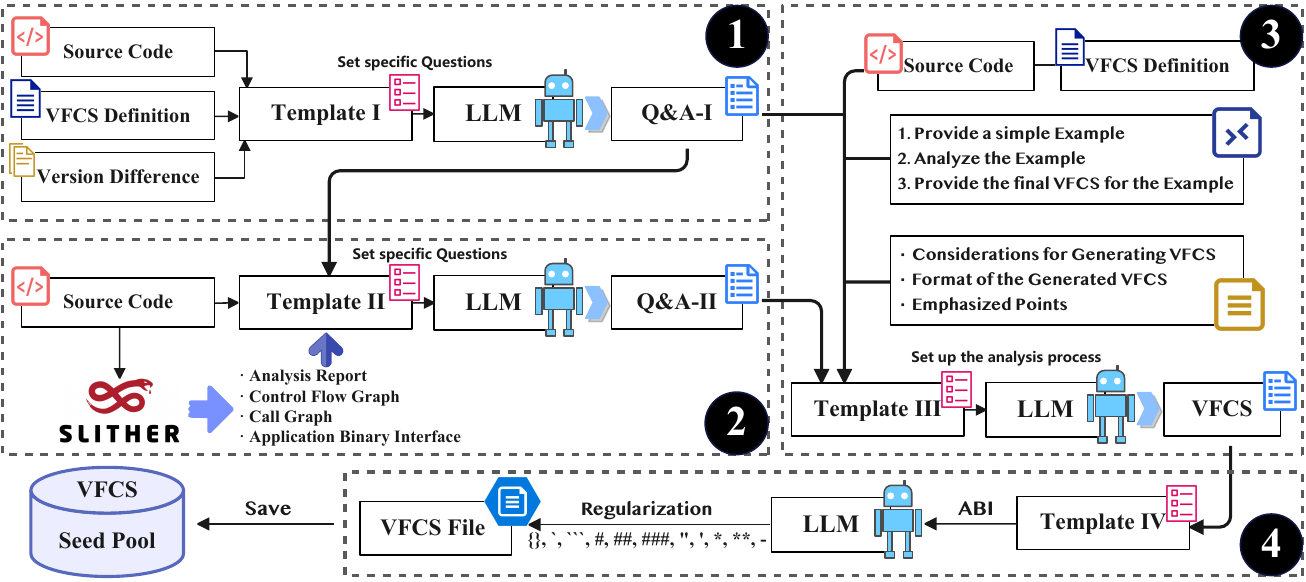}
    \vspace{-15pt}
    \caption{\small Chain-guided LLM Sequence Generation Process}
    \label{fig:chain-guided}
    \vspace{-17pt}
\end{figure}

Inspired by their methodology, we utilize LLMs to construct a chain-guided analysis procedure as shown in Figure~\ref{fig:chain-guided}, which allows us to logically generate candidate VFCS.
Next, we will describe each step in detail, and we have added a regularization module at the end to improve the availability of candidate VFCS, the effects of which are discussed in Section~\ref{sec:vfcs-llm}.

\PP{Step 1: High-risk Area Parsing} The LLM processes the contract source code via Template-I to extract critical functions and variables, while identifying potential offsetting function call relationships. It utilizes the concept of VFCS, which is provided as part of the input, to guide this analysis. Additionally, the model references a document detailing impossible vulnerability types for different contract versions to reduce model hallucinations. For example, in Figure~\ref{fig:crowdsale}, it identifies \code{invest} and \code{withdraw} as high-risk functions in the \texttt{Fundraiser} contract, and recognizes the offsetting relationship between \code{invest} and \code{refund}, highlighting the need to carefully examine their calling patterns.

\PP{Step 2: Static Validation} By integrating the contract source code with the static analysis tool \textit{Slither}, the LLM extracts precise function call information and control flow data. Using static analysis to enumerate function call combinations, the LLM validates the high-risk area parsing results from Step 1 and identifies potential function combinations that could trigger state changes or introduce vulnerability risks. For example, through enumerated call relationships, it identifies that a call sequence [\code{invest} $\to$ \code{invest}] under specific parameters would alter the \code{phase} variable state, effectively linking critical state variable changes to consecutive \code{invest} calls.

\PP{Step 3: Candidate VFCS Generation} Building upon the results of Step 1\&2, the LLM re-examines the source code and demonstrates the process of deriving VFCS from a sample contract using the gathered analytical and validation information. It outlines key points that need emphasis in VFCS generation, ensuring alignment with the ideal VFCS concept throughout. For example, by integrating reports from Step 1\&2, the LLM identifies [\code{invest} $\to$ \code{invest} $\to$ \code{withdraw}] as a VFCS through the rationale that first \code{invest} sets \code{invested>0}, second \code{invest} triggers \code{phase=1}, and \code{withdraw} triggers the vulnerability, representing the candidate VFCS to be generated.

\PP{Step 4: Regularization} In the final step, the LLM generates an ABI-compatible VFCS JSON file based on the ABI and the candidate VFCS. Automated scripts clean the file, ensuring it complies with ABI format and JSON standards. The final verified files are stored in the seed pool for fuzzing, ready for automated vulnerability testing.

The chain-guided approach analyzes contract-specific logic and inter-function interactions to generate candidate VFCSs that target vulnerability detection.
Unlike prior LLM-driven methods that focus on code prioritization and input generation, our approach generates VFCS candidates that are logically aware of state transitions within contracts (for more details, see Section~\ref{sec:related-work}).

\vspace{-2pt}
\subsection{LLM-Guided Iterative Fuzzing Process}
\label{sec:llm-guidedIFP}

In practical applications, contract state transitions are often obscured by complex parameter constraints, requiring specific transaction sequences to reach branches controlled by intricate variables. Existing fuzzers encounter two main limitations: (1) they either fail to alter the transaction sequence, repeating tests on the same paths, thereby hindering deep branching exploration, or (2) they modify the sequence within rigid predefined rules, leading to repetitive fragments and limited access to targeted branches. These issues create a bottleneck in fuzzing, especially when navigating deep, nested conditions where critical branching points reside.

\begin{figure} [t]
    \vspace{-3pt}
    \centering
    \includegraphics[width=1\linewidth]{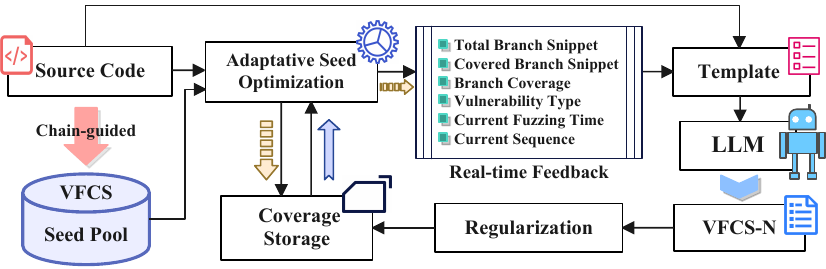}
    \Description{}
    \vspace{-18pt}
    \caption{\small Feedback-Driven LLM-guided Iterative Fuzzing Process}
    \label{fig:iterationProcess}
    \vspace{-15pt}
\end{figure}

To tackle this challenge, we propose a framework based on an LLM-guided iterative fuzzing process, designed to dynamically refine sequence generation through iterative feedback. 
The process starts by selecting a candidate VFCS from the \emph{seed pool}, generated through a chain-guided procedure (see Section~\ref{sec:vfcs-generation}), and using it in fuzzing to extract real-time feedback specific to the contract. This feedback includes key data such as \emph{Total Branch Snippet}, \emph{Covered Branch Snippet}, \emph{Branch Coverage}, identified \emph{Vulnerability Type}, \emph{Current Fuzzing Time}, and the \emph{Current Sequence}. The feedback is then embedded into a predefined template, which prompts the LLM to generate new contract-focused sequences aimed at reaching deeper, unexplored branches. After each fuzzing iteration, the real-time feedback and coverage data are stored as part of the \emph{Coverage Storage}. 
Once the LLM produces a new candidate VFCS, the saved coverage is reloaded, and the next fuzzing round begins. This iterative process creates a continuous loop of coverage storage, feedback generation, and sequence optimization. Figure~\ref{fig:iterationProcess} presents a detailed visualization of each step in this iterative process, illustrating how sequence generation evolves based on both contract properties and fuzzing outcomes. Throughout the fuzzing process, we apply branch distance-based seed mutation to guide the direction of seed mutation, continuously adapting and optimizing as reflected in the \emph{Adaptive Seed Optimization} step in the diagram.

By integrating the LLM-guided approach, our method expands the sequence range and introduces logical understanding to the mutation process. 
This allows the fuzzer to generate more meaningful mutations, crucial for complex smart contracts with deep branching conditions that simple random mutations often overlook.

\PP{Adaptive Seed Optimization} For function parameter generation, \ToolName{} follows the same proven strategies as state-of-the-art tools like IR-Fuzz and MuFuzz, using randomly generated seeds with branch distance-guided mutations towards unexplored paths. Each iteration refines the mutations, optimizing seeds for deeper exploration of complex branching structures. This adaptive strategy enhances coverage of critical code paths, ensuring the fuzzer efficiently uncovers hidden vulnerabilities and improves overall effectiveness in identifying potential security flaws.

\section{Experiment}
\label{sec:experiments}

In this section, we conduct extensive experiments to evaluate \ToolName{} with the aim of answering the following research questions:

\begin{itemize}[wide=5pt, topsep=2pt, itemsep=2pt, leftmargin=\dimexpr\labelwidth + 7.2 \labelsep\relax]
    \item[\textbf{RQ1.}] Does \ToolName{} demonstrate superior performance compared to state-of-the-art tools? 
    \item[\textbf{RQ2.}] What are the contributions of different components within \ToolName{}?
    \item[\textbf{RQ3.}] What is the impact of different LLMs on VFCS performance, and which LLM performs best?
\end{itemize}
In the following, we first introduce the experiment settings, followed by answering the above questions one by one. 
Additionally, we provide a case study to illustrate our workflow.

\subsection{Experiment Settings}

\subsubsection{Implementation} 
\ToolName{} is composed of over 2,000 lines of Python code, 4,800 words of prompts and 6,000 lines of C++ code. We implement \ToolName{} on the basis of IR-Fuzz~\cite{liu2023rethinking} (a state-of-the-art smart contract fuzzer).
For each component, our templates follow a unified structure: (1) Source \& intermediate results, (2) Analysis methodology, (3) Analysis objectives and outputs. Each step uses specific prompts to guide LLM analysis, including High-risk Area Parsing (Template-I), Static Validation (Template-II), Candidate VFCS Generation (Template-III), and Iterative Refinement (Template-Sec~\ref{sec:llm-guidedIFP}). All detailed templates are available at: \textcolor{blue}{\url{https://github.com/fuzzersc/vulnerabilities/tree/main/templates}}.
Our \ToolName{} framework, including all prompt templates and related tools, is open-sourced on GitHub.

\begin{table}[h]
\vspace{-5pt}
\centering
\caption{\small Statistical Overview of Evaluated Datasets.}
\vspace{-10pt}
\label{tab:statistical}
{\fontsize{8.5}{10.5}\selectfont % 
\setlength{\tabcolsep}{3pt}% 
\renewcommand{\arraystretch}{0.9}% 
\begin{tabular}{cccc}
\toprule
\textbf{Avg. Metric in Dataset (Count)} & \textbf{D1 (202)} & \textbf{D2 (143)} & \textbf{D3 (145)} \\ 
\midrule
Contract Size (KB)      & 3.49  & 3.44  & 7.58  \\
Functions per Contract  & 8.31  & 6.44  & 22.46 \\
Parameters per Function & 1.33  & 1.17  & 1.73  \\
State Variables per Contract         & 5.23  & 4.99  & 6.74  \\
Lines of Code per Contract           & 92.78 & 84.72 & 192.49 \\ 
\bottomrule
\end{tabular}
}
\vspace{-10pt}
\end{table}

\vspace{-3pt}
\subsubsection{Dataset}
Following prior work~\cite{torres2021confuzzius, ase2020smartbugs, liu2023rethinking}, we extend existing datasets, resulting in three distinct datasets:
\textbf{D1}: A mixed dataset of 202 medium and large contract projects, randomly selected from prior studies for diversity and cost control, as multiple LLMs, including high-cost, closed-source models, are used in our testing.
\textbf{D2}: A dataset of 143 contract projects with labeled vulnerabilities from previous research~\cite{ase2020smartbugs}. This collection provides a benchmark for evaluating the vulnerability detection capabilities of our method.
\textbf{D3}: A data set of 146 large and complex contract projects that we manually scraped from Etherscan~\cite{EtherscanWebsite}. These on-chain contracts add a real-world element to our analysis, enabling us to test our approach on live, previously unseen contracts.
Table~\ref{tab:statistical} presents the statistical overview of these three datasets. Dataset D3 represents the largest contracts, with 7.58KB average size and 22.46 functions per contract.
Given that contracts include extensive comments (helpful for LLM understanding), we classify contract-scale using source file size rather than bytecode instructions: Medium-scale (1-6KB), Large-scale ($>$6KB). 
We focus on medium- and large-scale contracts, as existing tools handle <1KB contracts effectively~\cite{jiang2018contractfuzzer,grieco2020echidna,ase2020smartbugs},
and the performance on <1KB contracts is further discussed in Section ~\ref{sec:discussion}.
This classification aligns with prior work~\cite{he2019learning,liu2023rethinking,qian2024mufuzz} that employs similar scale distinctions based on instruction count, typically using 3,000 instructions as the threshold for large-scale contracts, as shown in Table~\ref{tab:token_size}.

\begin{table}[h]
\vspace{-5pt}
\centering
\caption{\small Overview of selected LLMs and their advantages. In Type column, O: Open-Source; C: Closed-Source. In Usage column, LD: Local Deployment.}
\vspace{-10pt}
\label{tab:llm-choosing}
{\fontsize{4.8}{6.2}\selectfont 
\setlength{\tabcolsep}{3pt}%
\renewcommand{\arraystretch}{0.9}%
\setlength{\heavyrulewidth}{0.08em}    % Width of thick lines at the top and bottom
\setlength{\lightrulewidth}{0.05em}    % The width of the thin line in the middle
\setlength{\belowrulesep}{0.6ex}       % Spacing below the line
\setlength{\aboverulesep}{0.6ex}       % Spacing on the line
\begin{tabular}{l c l l c}  %
    \toprule
    \textbf{LLMs} & \textbf{Type} & \textbf{Available URL} & \textbf{Advantages} & \textbf{Usage} \\
    \midrule
    \textbf{Qwen2}-7B         & \textbf{O}   & \url{https://ollama.com/library/qwen}               & Widely Used lightweight LLM in Chinese    & LD \\
    \textbf{Llama3.1}-8B      & \textbf{O}   & \url{https://ollama.com/library/llama3.1}           & Widely Used lightweight LLM               & LD \\
    \textbf{Llama3.1}-405B    & \textbf{O}   & \url{https://ollama.com/library/llama3.1}           & Strongest Open-source LLM                 & API              \\
    \textbf{GPT 3.5 Turbo}    & \textbf{C} & \url{https://platform.openai.com/} & Long Time Used and Cost-Effective LLM     & API              \\
    \textbf{GPT 4o-mini}      & \textbf{C} & \url{https://platform.openai.com/}   & Widely Used, Highly Cost-Effective LLM & API              \\
    \textbf{Claude3.5 Sonnet} & \textbf{C} & \url{https://docs.anthropic.com/}                   & Strongest Closed-source LLM currently     & API              \\
    \bottomrule
\end{tabular}%
}
\vspace{-10pt}
\end{table}

\subsubsection{Baseline}
To compare the performance and effectiveness of \ToolName{}, we consider both the LLM and the Fuzzer aspects.

\PP{LLM}
Table~\ref{tab:llm-choosing} compares the LLMs selected, emphasizing their characteristics and selection rationale. Our diverse choices cover various use cases, including lightweight open-source models like Qwen2-7B and LLaMA3.1-8B for local deployment, and the LLaMA3.1-405B as the strongest open-source model. To ensure cost-effectiveness, we also include the widely used GPT 4o-mini and GPT 3.5 Turbo for API deployment. Lastly, we selected Claude 3.5 Sonnet, the most powerful closed-source model, to explore peak performance. This varied selection allows for a thorough evaluation of our framework's usability and performance across different model types.

\PP{Fuzzer}
We include sFuzz~\cite{nguyen2020sfuzz}, IR-Fuzz~\cite{liu2023rethinking}, and MuFuzz~\cite{qian2024mufuzz} in our comparisons for two reasons: (1) Their performance has been experimentally proven to surpass other smart contract fuzzing tools, such as ContractFuzzer~\cite{jiang2018contractfuzzer}, Echidna~\cite{grieco2020echidna}, ContraMaster~\cite{wang2019vultron,choi2021smartian}, ILF~\cite{he2019learning}, and Confuzzius~\cite{torres2021confuzzius}; and (2) these tools are commonly used as benchmarks in current research. Currently, MuFuzz is considered the state-of-the-art fuzzer. 
We do not compare with LLM4Fuzz~\cite{shou2024llm4fuzz} as it lacks open-source availability and accessible artifacts.
A technical comparisons are provided in Section~\ref{sec:related-work}.
We compare \ToolName{} against these fuzzers in terms of branch coverage and vulnerability detection. For each fuzzing process, we perform fuzzing with all three generated VFCS candidates and take the average results for comparison. When 60s elapses without discovering new paths, the fuzzing process transitions from the initial VFCS to an LLM-driven iterative process, designed to explore more efficient paths.

\PP{Device}
We conduct our evaluations on an Ubuntu 20.04 system equipped with an Intel(R) Xeon(R) Platinum 8255C CPU (12 virtual cores, 2.50GHz), 40GB RAM, and a 10GB RTX 3080 GPU.

\subsection{Performance Evaluation (RQ1)}
We employ Claude 3.5 Sonnet as the LLM in our experiments, which has been proven to be the strongest among selected LLMs in Section~\ref{sec:vfcs-llm}. Subsequently, we compare the performance of \ToolName{}, which is implemented following the workflow shown in Figure~\ref{fig:overview}, with several other fuzzers. The comparison is based on two metrics: \emph{vulnerability detection} and
\emph{branch coverage}.

\begin{table*}[t]
\centering
\caption{\small Results show by format \textbf{IR-Fuzz/MuFuzz/\ToolName{}}, respectively. \ToolName{} against state-of-the-art fuzzers in Vulnerability Detection of Different Dataset. In Oracle setting, GL: Gasless, IO: Integer Over-/under-flow, RE: Reentrancy, UC: Unchecked Call, BN: Block Number Dependency, TP: Timestamp Dependency, DG: Dangerous Delegatecall, UE: Unexpected Ether or Ether Frozen. In ET row, it means Each Type.}
\label{tab:vul-sota}
\vspace{-8pt}
\includegraphics[width=\textwidth]{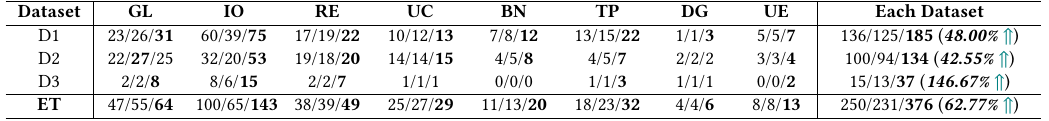}
\vspace{-18pt}
\end{table*}

\vspace{-1pt}
\subsubsection{Vulnerability Detection}

We conduct vulnerability detection in datasets D1, D2, and D3 using \ToolName{}, in conjunction with state-of-the-art fuzzing tools. Notably, we adopt a strict oracle standard that is consistent with prior works such as IR-Fuzz and MuFuzz. 
Given that
fuzzing typically reveals true vulnerabilities and that previous studies have thoroughly analyzed false positives and false negatives under this oracle standard~\cite{liu2023rethinking, qian2024mufuzz}, we focus on carefully evaluating \ToolName{}'s effectiveness in identifying vulnerabilities across different datasets and understanding their distribution more comprehensively.

The results, presented in Table~\ref{tab:vul-sota} in the format IR-Fuzz/MuFuzz/\ToolName{}, demonstrate that \ToolName{} consistently outperforms state-of-the-art fuzzers across various datasets and vulnerability types. Specifically, \ToolName{} detects 48.00\%, 42.55\%, and 146.67\% more vulnerabilities than MuFuzz on D1, D2, and D3, respectively, with an overall average improvement of 62.77\%. These results highlight the substantial advantages of \ToolName{} over existing tools in various scenarios, particularly when analyzing new, large-scale, and complex real-world contract projects on the blockchain (dataset D3). 
As contract structures become increasingly complex, traditional approaches that rely solely on syntactic patterns struggle to effectively uncover meaningful attack sequences.
We attribute the superior performance of \ToolName{} to its logic-driven approach, which extracts VFCS to optimize state transitions and takes advantage of real-time feedback to guide the LLM in exploring contract logic and simulating attacks that exploit the unique logic of each contract. Encouragingly, we have discovered 37 previously unknown vulnerabilities from 19 newly contract projects, whereas other state-of-the-art tools could only identify 13 and 15 vulnerabilities, respectively. The details are put in anonymous URL: \textcolor{UrlColor}{\url{https://github.com/fuzzersc/vulnerabilities}}.

\begin{figure}[b]
    \vspace{-10pt}
    \centering
    \begin{subfigure}[b]{0.235\textwidth}
        \centering
        \vspace{-5pt}
\includegraphics[width=\textwidth]{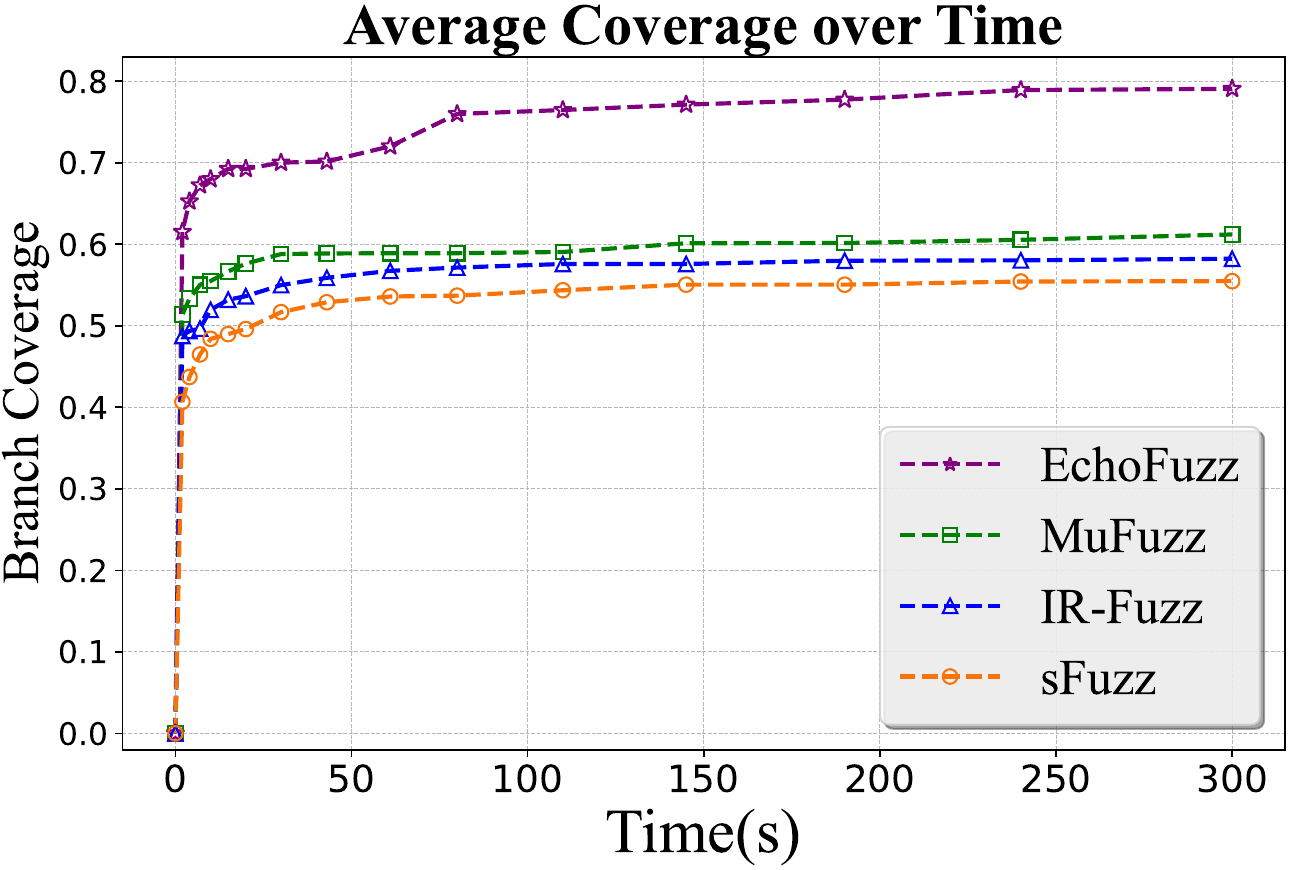}
        \vspace{-15pt}
        \caption{\small On Medium contracts}
        \label{fig:echo-sota-a}
    \end{subfigure}
    \begin{subfigure}[b]{0.235\textwidth}
        \centering
    \vspace{-5pt}
\includegraphics[width=\textwidth]{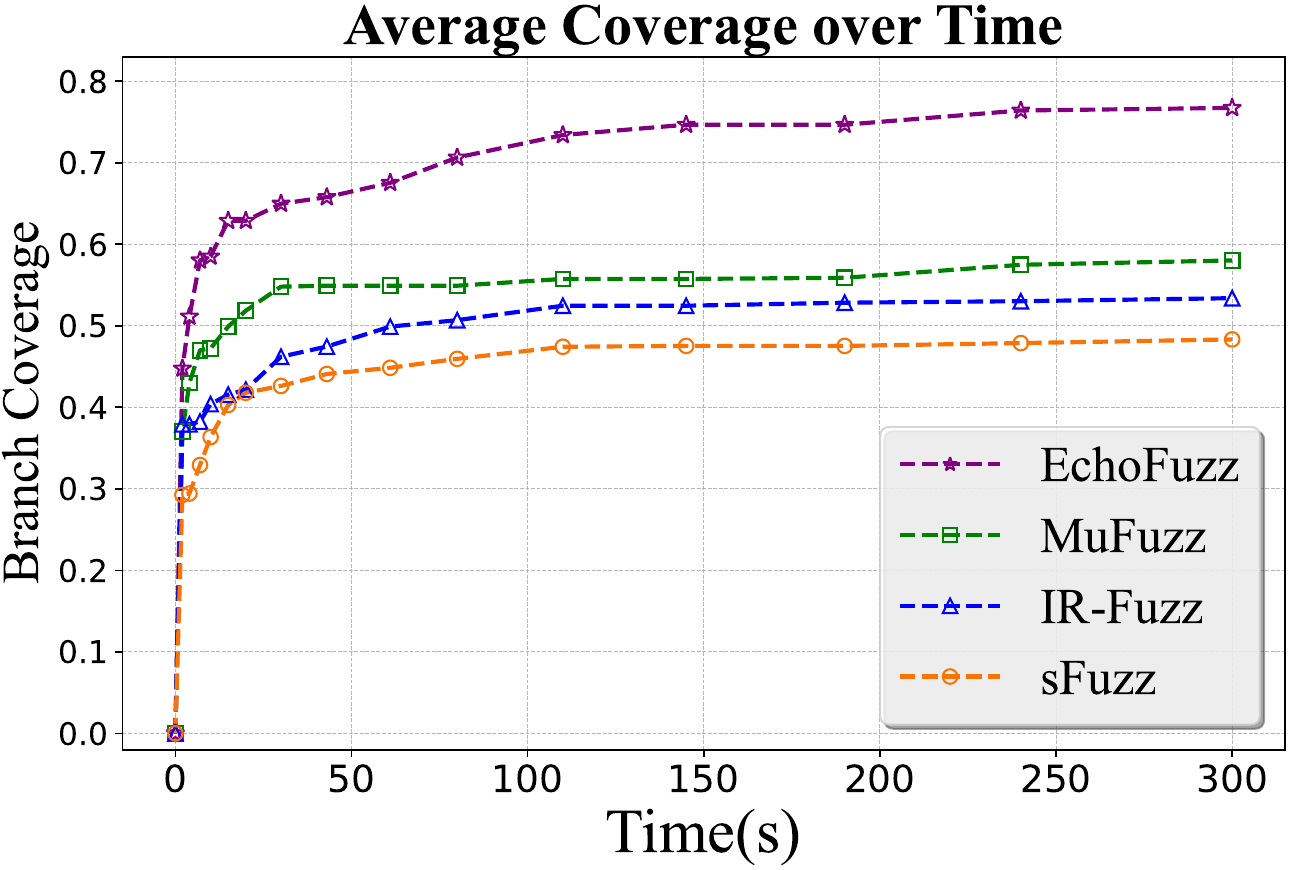}
        \vspace{-15pt}
        \caption{\small On Large contracts}
        \label{fig:echo-sota-b}
    \end{subfigure}
    \vspace{-22pt}
    \caption{\small Branch coverage comparison between \ToolName{} and the state-of-the-art fuzzers.}
    \label{fig:echofuzz-sota}
    \vspace{-3pt}
\end{figure}

\vspace{1.5pt}
\PP{False Negative Analysis} We conduct a detailed review of the two false negatives of \ToolName{} compared to MuFuzz on GL, shown in Table~\ref{tab:vul-sota}. The review reveals that the issue stems from insufficient cumulative iteration times of the seeds and the limitations of seed exploration in \ToolName{}. In the context of simple contracts, the function call sequences are relatively fixed. Compared to \ToolName{}, MuFuzz follows rule-based mutation, can maintain a simple sequence while spending more time mutating seeds, allowing seeds to reach and vary in a wider range, and thus reach deeper branches. 
This represents a limitation of our approach, which we discuss in more detail in the seed mutation part of Section~\ref{sec:discussion}.

\vspace{-3pt}
\subsubsection{Branch Coverage}

We evaluate branch coverage of \ToolName{}, sFuzz, IR-Fuzz, and MuFuzz on medium and large contracts in D1. \ToolName{} consistently outperforms all others, as shown in Figure~\ref{fig:echofuzz-sota}.
On medium contracts, \ToolName{} achieves 79.1\% coverage, surpassing sFuzz, IR-Fuzz, and MuFuzz by 38.40\%, 35.80\%, and 29.19\%, respectively (Figure~\ref{fig:echo-sota-a}).
On large contracts (Figure~\ref{fig:echo-sota-b}), all tools show lower coverage, but \ToolName{}'s decline is minimal (2.4\%). The declines of others are 3.76$\times$, 2.07$\times$, and 1.36$\times$ greater. \ToolName{} still achieves 76.70\% coverage, outperforming others by 58.75\%, 43.71\%, and 32.24\%, demonstrating superior robustness.
\ToolName{} outperforms other fuzzers on large contracts, achieving higher coverage rates in less time. This is due to its chain-guided VFCS generation, which enhances testcase quality beyond that of fuzzers relying on Random and Read-Write dependency strategies.

Furthermore, \ToolName{} continues to increase coverage effectively, while other fuzzers typically see their performance gains rapidly diminish after 60s, quickly hitting fuzzing bottlenecks that are hard to overcome.
Beyond this, limited state space sharply reduces effectiveness, yielding diminishing returns with minimal further improvement.
Further analysis reveals that the coverage improvement is more substantial for large contracts.
\ToolName{} achieves a 10.02\% improvement on medium contracts, while other fuzzers only improve by around 3\%, demonstrating a 3$\times$ performance boost. This suggests that, after a period without discovering new paths, the fuzzing process transitions from the initial VFCS to an LLM-driven iterative approach. In this phase, fuzzers are no longer limited by their fixed state space. Even MuFuzz, which employs sequence-aware mutation strategies, struggles to explore new branches effectively. In contrast, \ToolName{} continues to demonstrate impactful mutations. 
This is due to LLM-guided VFCS generation: leveraging logic insights and real-time feedback, the LLM analyzes covered code and generates targeted tests for unexplored branches, enhancing mutation effectiveness.

{
\vspace{5pt}
\centering
\includegraphics[width=0.48\textwidth]{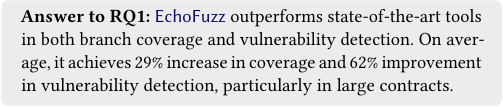}
\vspace{-8pt}
}

\subsection{Ablation Study (RQ2)}

By default, \ToolName{} uses a chain-guided LLM to generate high-quality candidate VFCSs for fuzzing. We are curious about how each component within the chain-guided procedure contributes to the overall performance. Furthermore, it is interesting to observe the effects of removing the entire chain-guided strategy. We seek to understand how the two core LLM-utilizing approaches compare against pure static analysis to isolate the LLM's contribution. Additionally, \ToolName{} uses real-time fuzzing feedback to guide the LLM in generating the new VFCS for the fuzzing process. We are also curious to see how much this method contributes to the overall performance of \ToolName{}. Next, we conduct experiments on each of these key components separately.

\begin{figure}[b]
    \vspace{-10pt}
    \centering
    \begin{subfigure}[b]{0.235\textwidth}
        \centering
        \vspace{-5pt}
\includegraphics[width=\textwidth]{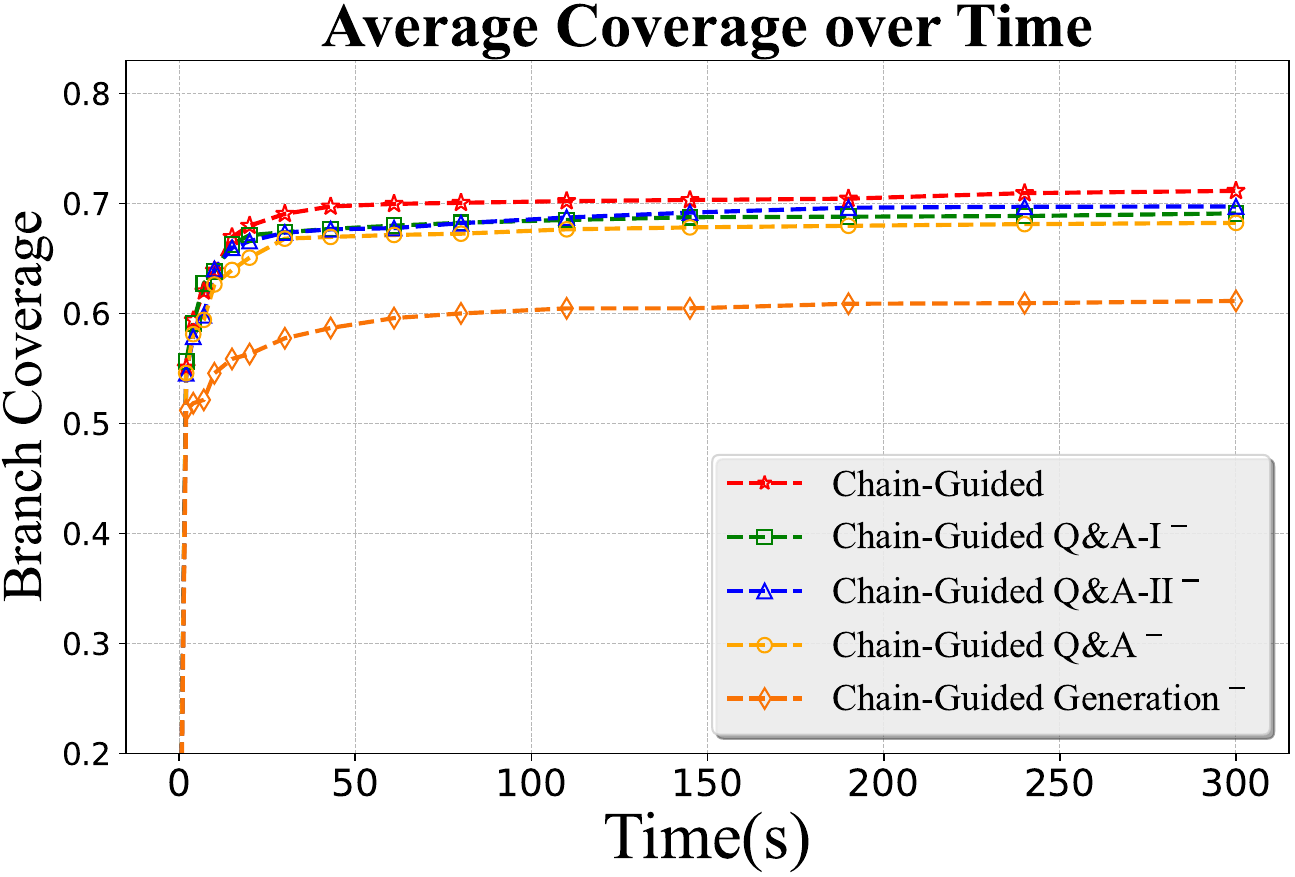}
        \vspace{-15pt}
        \caption{\small On Medium contracts}
        \label{fig:cg-ablation-medium}
    \end{subfigure}
    \begin{subfigure}[b]{0.235\textwidth}
        \centering
    \vspace{-5pt}
\includegraphics[width=\textwidth]{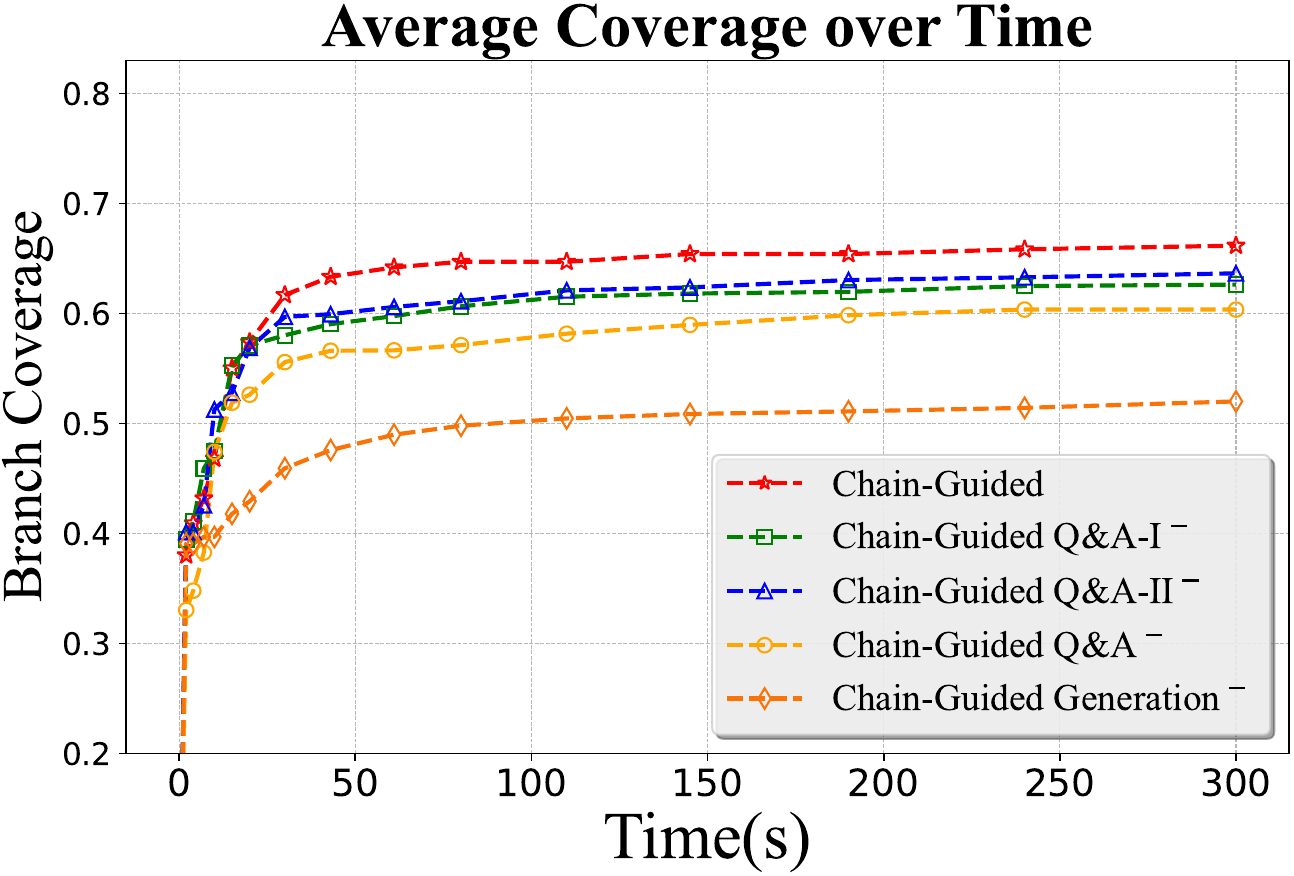}
        \vspace{-15pt}
        \caption{\small On Large contracts}
        \label{fig:cg-ablation-large}
    \end{subfigure}
    \vspace{-22pt}
    \caption{\small Ablation study of chain-guided LLM procedure in \ToolName{} on branch coverage.}
    \label{fig:cg-ablation}
    \vspace{-5pt}
\end{figure}

\vspace{-2pt}
\subsubsection{Chain-guided Framework Component Analysis}
We conduct ablation experiments by removing components: Q\&A-I (High-risk Area Parsing), Q\&A-II (Static Validation), both Q\&A steps (Q\&A$^{-}$), and Candidate VFCS Generation (Generation$^{-}$).
Figure~\ref{fig:cg-ablation} presents the branch coverage performance comparison. The results shows that removing any component degrades performance, with escalating impact on larger contracts. Individual Q\&A step removal causes 2.96\%-5.66\% coverage loss, while their collective removal (Q\&A$^{-}$) yields 4.26\%-9.63\% degradation, demonstrating synergistic effects. Most critically, Generation$^{-}$ causes severe 16.37\%-26.41\% degradation, as this component synthesizes Q\&A insights into actionable VFCS candidates—without it, the framework reduces to dependency methods like existing tools.
These results validate the effectiveness of each component, particularly for large-scale contracts where multi-step reasoning proves essential.

\vspace{-2pt}
\subsubsection{Study on Chain-guided Generating Initial VFCS}
We replace chain-guided LLM process in \ToolName{} with random sequence generation, creating \textbf{EchoFuzz-WCG}(``\textbf{W}ithout \textbf{C}hain-guided \textbf{G}eneration'').
This variant samples functions randomly based on their distribution in the contract code.
As Figure~\ref{fig:ablation} shows, EchoFuzz-WCG initially lags behind EchoFuzz-SA in the first 60s due to the absence of targeted guidance. However, it later surpasses EchoFuzz-SA by leveraging LLM-guided iterations.
In contrast, the full chain-guided version (EchoFuzz-WIP) outperforms EchoFuzz-SA immediately, achieving 28\% higher coverage in the first few seconds.e
At 300s, \ToolName{} achieves 6.98\% higher coverage than EchoFuzz-WCG and identifies 24.09\% more vulnerabilities on average.
The chain-guided LLM procedure provides immediate effectiveness through multi-step reasoning and high-quality initial VFCS generation, removing random exploration phases and enabling targeted vulnerability discovery from the start of fuzzing.

\vspace{-2pt}
\subsubsection{Study on LLM-guided Iteration Process}
\begin{figure}[b]
    \vspace{-20pt}
    \centering
    \begin{subfigure}[b]{0.235\textwidth}
        \centering
        \includegraphics[width=\textwidth]{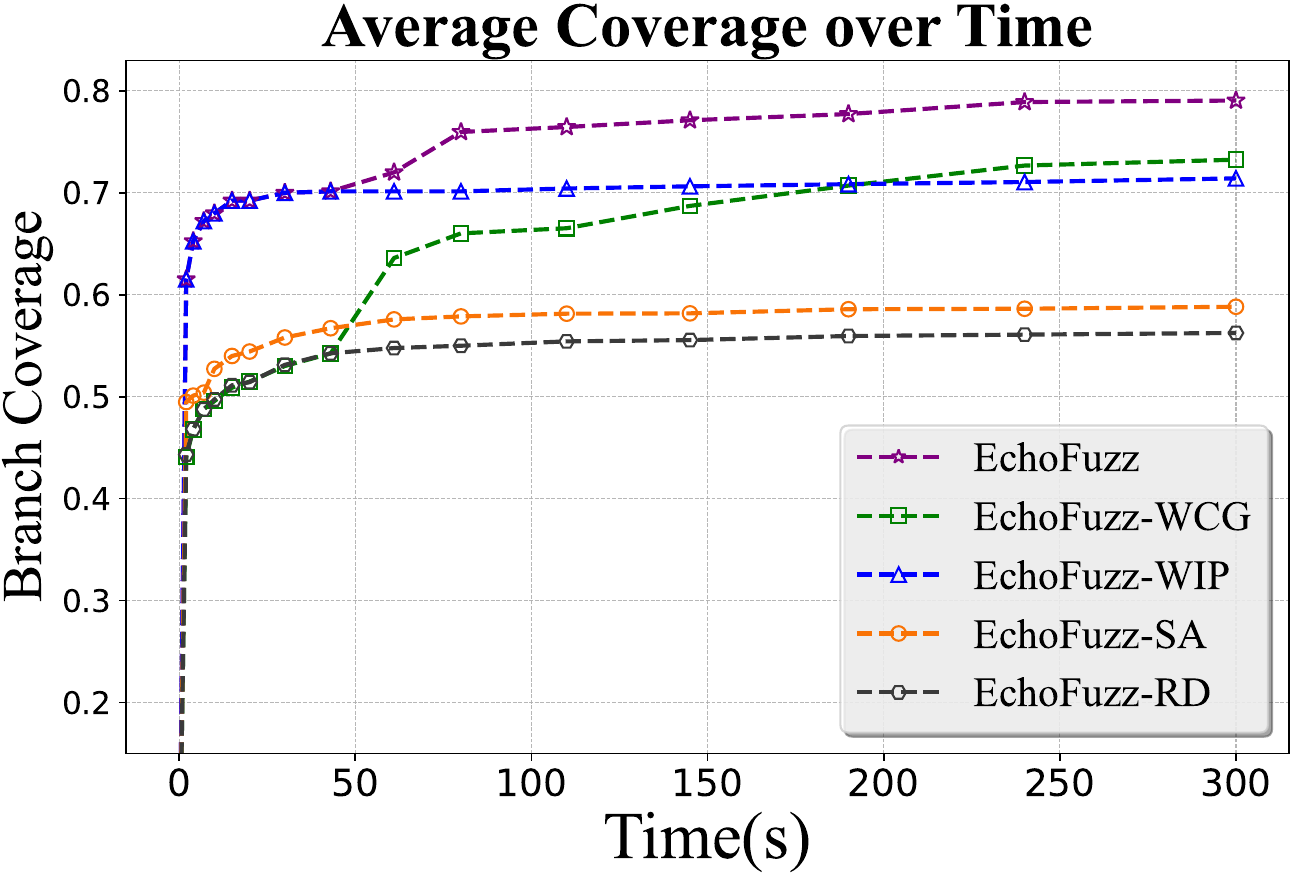}
        \vspace{-15pt}
        \caption{\small On Medium contracts}
        \label{fig:echo-ablation-b}
    \end{subfigure}
    \begin{subfigure}[b]{0.235\textwidth}
        \centering
        \includegraphics[width=\textwidth]{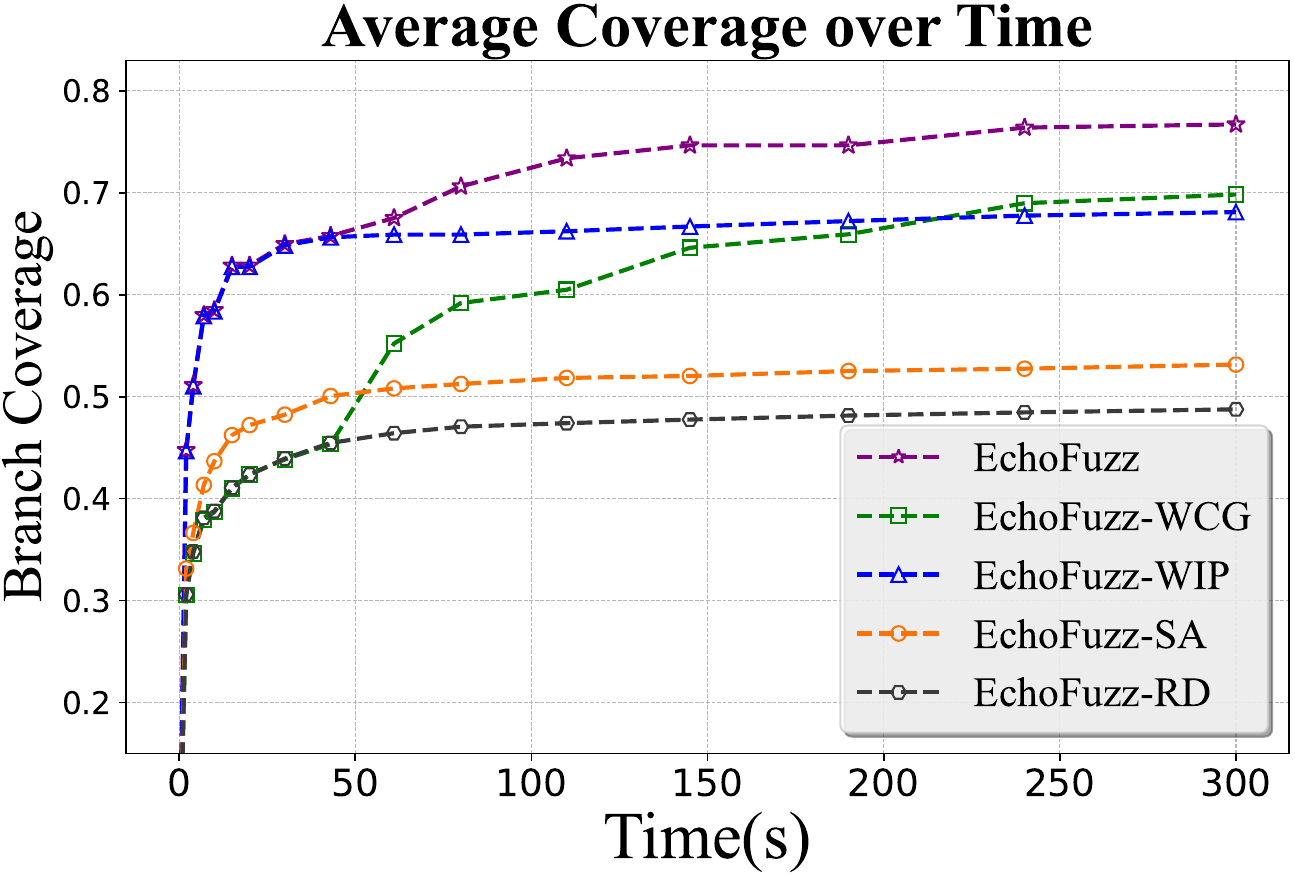}
        \vspace{-15pt}
        \caption{\small On Large contracts}
        \label{fig:echo-ablation-b}
    \end{subfigure}
    \vspace{-20pt}
    \caption{\small Ablation study of \ToolName{}.}
    \label{fig:ablation}
    \vspace{-0pt}
\end{figure}

We remove the LLM-guided iterative fuzzing process from \ToolName{}, conducting fuzzing with only the original VFCS without further modifications, resulting in \textbf{EchoFuzz-WIP} (``\textbf{W}ithout LLM-guided \textbf{I}teration \textbf{P}rocess''). This variant employs the chain-guided LLM procedure but lacks real-time feedback-driven sequence refinement.
As shown in Figure~\ref{fig:ablation} and Table~\ref{tab:vul-ablation}, EchoFuzz-WIP demonstrates immediate effectiveness through chain-guided generation, achieving 28\% higher coverage than EchoFuzz-SA in the first few seconds and significantly outperforming EchoFuzz-WCG in beginning. However, the complete \ToolName{} framework shows a 10.65\% coverage improvement over EchoFuzz-WIP at 300s. In vulnerability detection, particularly in complex contracts (D3), \ToolName{} identifies 54.17\% more vulnerabilities than EchoFuzz-WIP.
LLM-guided iteration process enables \ToolName{} to overcome the exploration bottleneck faced by other fuzzers between 60s-300s, achieving 22.06\% coverage improvement (vs. 2-5\% for other fuzzers). This nearly order-of-magnitude enhancement over existing tools is extremely encouraging.

\begin{table*}[t]
\centering
\caption{\small \ToolName{} each components in Vulnerability Detection of Different Dataset. Relevant oracle setting can see Table 3. Results show by format \textbf{EchoFuzz-RD/EchoFuzz-SA/EchoFuzz-WCG/EchoFuzz-WIP/\ToolName{}}, respectively.}
\label{tab:vul-ablation}
\vspace{-10pt}
\includegraphics[width=\textwidth]{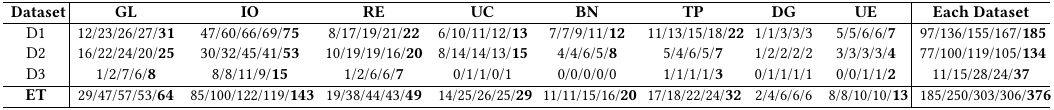}
\vspace{-21pt}
\end{table*}

\vspace{-5pt}
\subsubsection{LLM Contribution Analysis: Comparison with Non-LLM Baselines}
To isolate the LLM's contribution, we introduce two non-LLM baselines: \textbf{EchoFuzz-RD} (``\textbf{R}an\textbf{d}om'') and \textbf{EchoFuzz-SA} (``\textbf{S}tatic \textbf{A}nalysis only''), enabling direct comparison with our LLM-based approaches EchoFuzz-WIP and EchoFuzz-WCG.
As shown in Figure~\ref{fig:ablation} and Table~\ref{tab:vul-ablation}, both LLM approaches achieve substantial improvements over non-LLM baselines. In coverage, they demonstrate more than 23\% and 29\% improvements over EchoFuzz-SA and EchoFuzz-RD respectively. In vulnerability detection, both achieve over 21\% and 63\% improvements over the non-LLM baselines respectively. The complete \ToolName{} framework achieves 50.4\% and 103.2\% vulnerability detection improvements over EchoFuzz-SA and EchoFuzz-RD respectively, validating the significant contribution of LLM-based guidance in smart contract fuzzing.

{\vspace{5pt}
\centering
\includegraphics[width=0.48\textwidth]{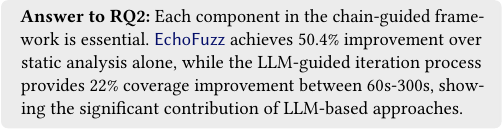}
\vspace{-8pt}}

\subsection{\fontsize{10.3}{10}\selectfont VFCS Performance with Different LLMs (RQ3)}
\label{sec:vfcs-llm}

We evaluate VFCS performance across different LLMs using two key metrics: \textit{branch coverage} and \textit{VFCS available ratio}. These metrics comprehensively assess both the quality and applicability of generated test cases across our contract dataset.

\begin{figure}[b]
    \vspace{-10pt}
    \centering
    \begin{subfigure}[b]{0.235\textwidth}
        \centering
        \includegraphics[width=\textwidth]{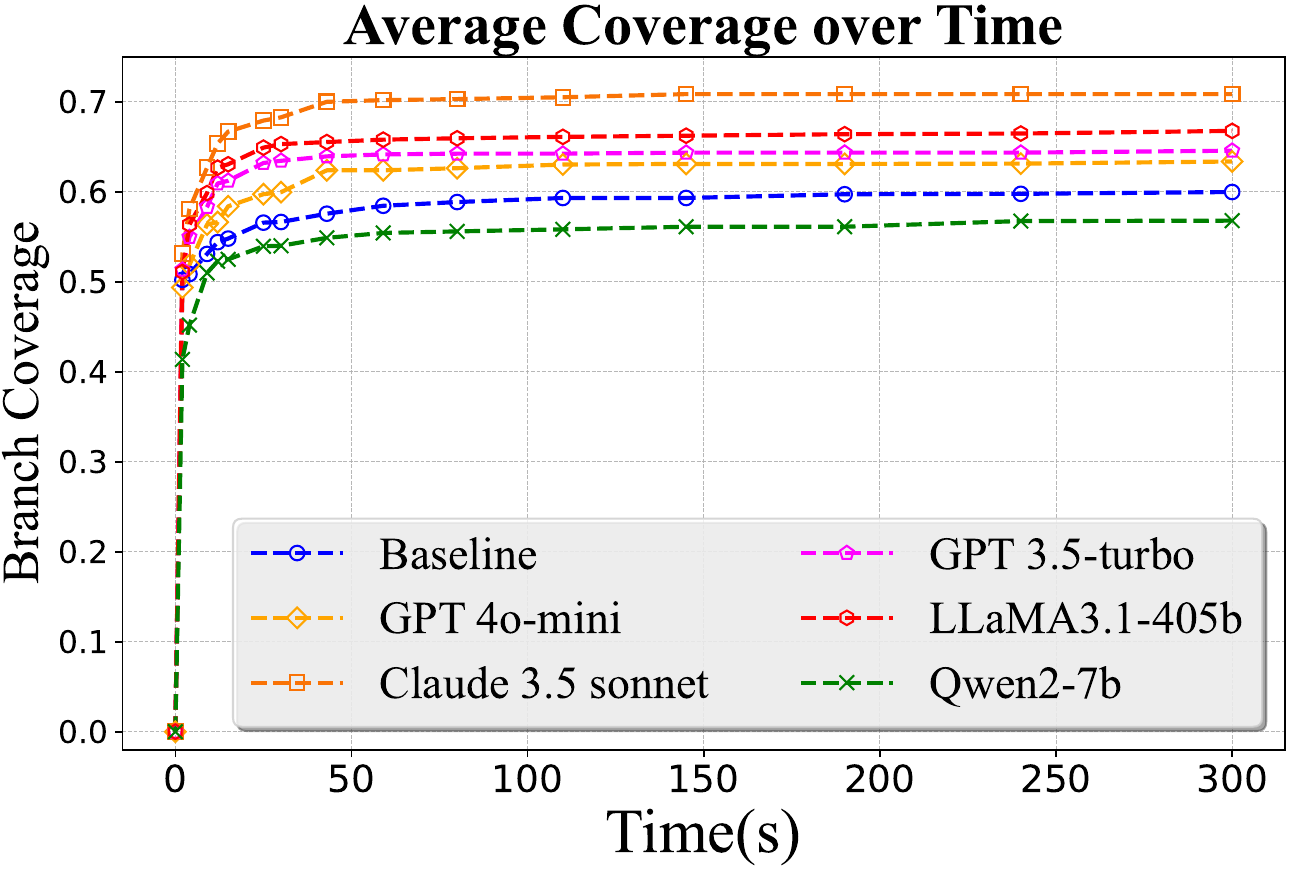}
        \vspace{-14pt}
        \caption{\small On Medium contracts}
        \label{fig:VFCS-diff_llm-a}
    \end{subfigure}
    \begin{subfigure}[b]{0.235\textwidth}
        \centering
        \includegraphics[width=\textwidth]{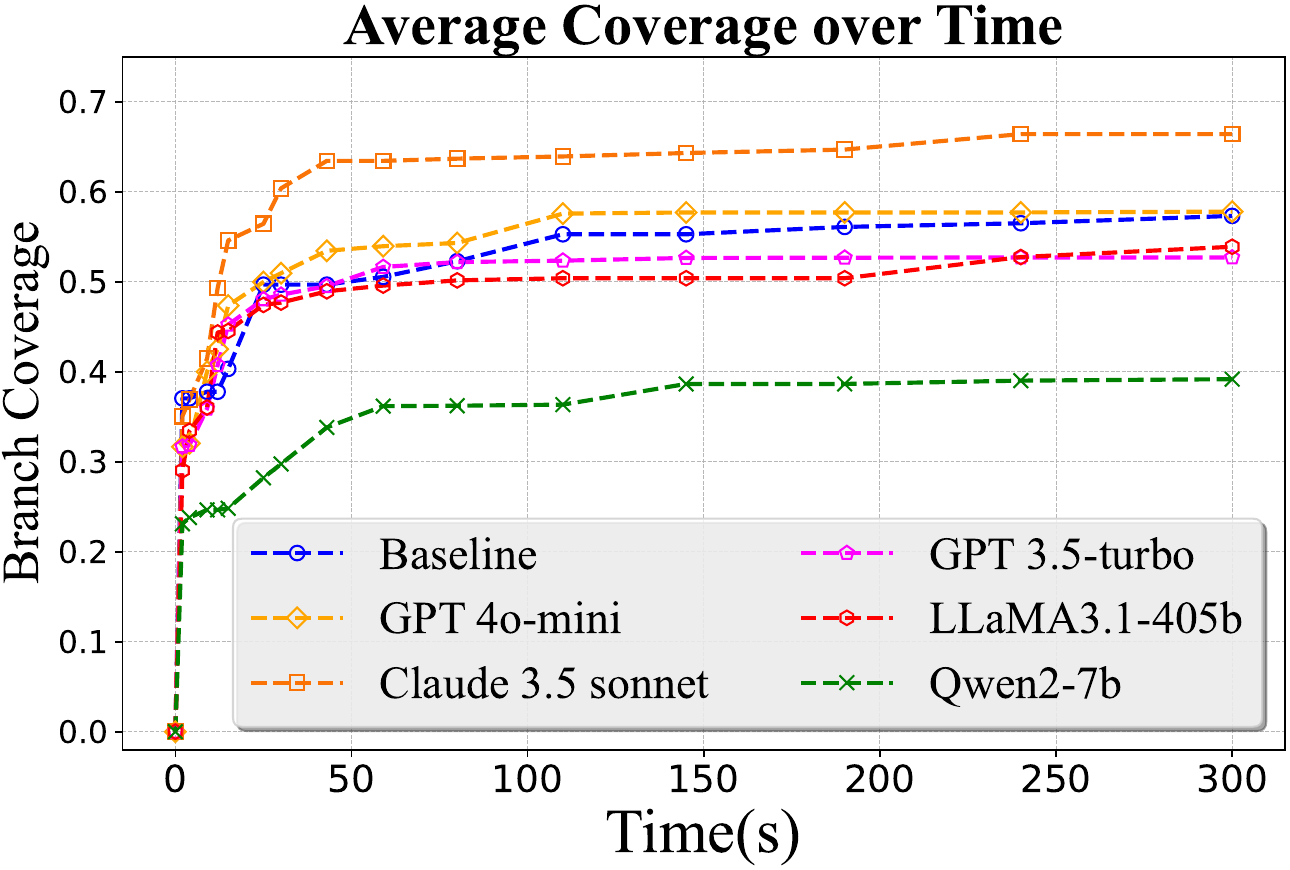}
        \vspace{-14pt}
        \caption{\small On Large contracts}
        \label{fig:VFCS-diff_llm-b}
    \end{subfigure}
    \vspace{-20pt}
    \caption{\small Different LLMs generation VFCS in Branch coverage.}
    \label{fig:VFCS-diffLLM}
    \vspace{-1pt}
\end{figure}

 \vspace{-2pt}
\subsubsection{Branch Coverage in Different LLMs}
We evaluate effectiveness of VFCS by measuring branch coverage in D1, comparing various LLMs with the baseline (IR-Fuzz) as shown in Figure~\ref{fig:VFCS-diffLLM}. Figure~\ref{fig:VFCS-diff_llm-a} shows results for medium contracts, and Figure~\ref{fig:VFCS-diff_llm-b} for large contracts, illustrating coverage trends over time.
Claude 3.5 Sonnet consistently outperforms both other models and the baseline once fuzzing stabilizes, regardless of contract size. On medium contracts, it achieves 70.84\% branch coverage, an impressive 18.14\% higher than the baseline. Other models, such as LLaMA3.1-405B, GPT 3.5-turbo, and GPT 4o-mini, show smaller gains of 11.32\%, 7.66\%, and 5.66\%, respectively, while Qwen2-7B underperforms with 5.31\% lower coverage than the baseline. On large contracts, Claude 3.5 Sonnet again leads, achieving a remarkable 15.85\% higher branch coverage than the baseline. GPT 4o-mini provides a modest improvement of 0.85\%, while LLaMA3.1-405B, GPT 3.5-turbo, and Qwen2-7B fall below the baseline by 5.98\%, 8.06\%, and 31.66\%, respectively.
For larger contracts, LLMs tend to prioritize vulnerability-triggering paths over broader branch coverage, limiting their effectiveness. This suggests the need for LLM-guided iteration combined with additional fuzzing to maximize coverage and uncover hidden paths.
These results highlight model size impact and closed-source models' advantages for VFCS.
Smaller models decline in effectiveness with larger contracts.
Claude 3.5 Sonnet demonstrates superior performance by achieving high branch coverage quickly, making it the preferred model for high-quality test case generation.

\vspace{-2pt}
\subsubsection{Available Ratio of VFCS Generated by Different LLMs}
To evaluate the availability of VFCS generated by different LLMs,
we designed our framework to generate three VFCS candidates per contract in dataset D1,
which contains 202 contracts, targeting a total of 606 VFCS files.
Initial experiments revealed that a significant portion of generated VFCS were unusable, leading to resource waste. By integrating a regularization module, our framework significantly improve VFCS available ratio. 
In Table~\ref{tab:vfcs_available}, “NoR” refers “No-use-Regularization,” with \textit{Quantity-NoR} as usable VFCS without regularization. \textit{Improve} reflects the availability gain from regularization, and \textit{Available Ratio} is usable VFCS relative to the target of 606.
As shown in Table~\ref{tab:vfcs_available}, LLaMA3.1-405B’s VFCS availability increases from 281 to 541, an increase of 92.5\%. Other LLMs also exhibit roughly a 50\% increase in availability with regularization.
In our framework, Claude 3.5 Sonnet achieves the highest availability ratio at 94.2\%, indicating that our approach is both practical and versatile across a wide range of contract types.
Similarly, GPT 3.5 Turbo and GPT 4o-mini both surpass the 90\% threshold, while LLaMA3.1 405B approaches 90\%, confirming the strong performance of larger models within our framework. In contrast, smaller models such as LLaMA3.1 8B achieve a notably lower availability ratio of 42.1\%, highlighting their limitations in domain-specific tasks without specialized fine-tuning. This points to a promising direction for future research: enhancing the performance of smaller, more accessible models via targeted training techniques.

\begin{table}[t]
\centering
\caption{\small Performance of Different LLMs on VFCS Available Ratio and Regularization Improvement.}
\label{tab:vfcs_available}
\vspace{-8pt}
{\fontsize{7.2}{9}\selectfont
\setlength{\tabcolsep}{3pt}
\renewcommand{\arraystretch}{0.88}% 
\begin{tabular}{lcccc}
\toprule
\multicolumn{1}{c}{\textbf{LLM Version}} & \textbf{Quantity-NoR} & \textbf{Quantity} & \textbf{Improve} & \textbf{Available Ratio} \\ 
\midrule
Qwen2-7B                                 & 263                   & 401               & 52.5\%           & 66.2\%                   \\
LLaMA3.1-8B                              & 182                   & 255               & 40.1\%           & 42.1\%                   \\
LLaMA3.1-405B                            & 281                   & 541               & \textbf{92.5\%}  & 89.3\%                   \\
GPT 3.5 Turbo                            & \textbf{385}          & 557               & 44.7\%           & 91.9\%                   \\
GPT 4o Mini                              & 345                   & 549               & 59.1\%           & 90.6\%                   \\
\textbf{Claude 3.5 Sonnet}               & 379                   & \textbf{571}      & 50.7\%           & \textbf{94.2\%}          \\ 
\bottomrule
\end{tabular}%
}
\vspace{-13pt}
\end{table}

{
\vspace{4pt}
\centering
\includegraphics[width=0.48\textwidth]{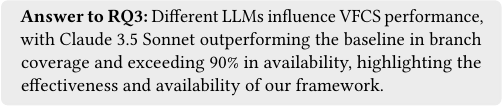}
\vspace{-12pt}
}

\subsection{Case Study}

To illustrate how \ToolName{} applies its two core procedures for targeted sequence generation in contract logic to detect complex vulnerabilities and overcome exploration bottlenecks, we present its workflow via a case study.
Figure~\ref{fig:crowdsale} presents the contract, with detailed explanations on how to trigger the bug provided in Section~\ref{sec:motivating-example}. 
Other tools fail to detect this bug as they 
overlook the impact of \code{refund} on the state variable \code{invested}, achieving only 50\% coverage.
In contrast, \ToolName{} generates a sequence capable of triggering the bug using chain-guided procedure (i.e., EchoCV in Figure~\ref{fig:intro-ab}, 
with detailed procedure 
on how to generate that VFCS in Section~\ref{sec:vfcs-generation}) to generate \emph{T1}: [\code{invest} $\to$ \code{invest} $\to$ \code{withdraw}]. 
After the first round of fuzzing, the bug is triggered at line 28.
At this point, the branch coverage reaches 75\%, as the \emph{T1} sequence does not call \code{refund}. In the LLM-guided iterative process, the LLM identifies that all branches of the \code{invest} and \code{withdraw} are covered, and a potential bug is found within \code{withdraw}. 
Based on real-time feedback, the LLM generates a new sequence to explore branches in \code{refund} 
(EchoCV-N in Figure~\ref{fig:intro-ab}): \emph{T2} = [\code{invest} $\to$ \code{refund}].
\emph{T2} incorporates the coverage information saved at the end of the first round of fuzzing to initiate the second round of fuzzing.
The second round of fuzzing ended in under a second, with branch coverage reaching 100\%, completing the process. This workflow demonstrates that \ToolName{} ensures in-depth exploration of contract vulnerabilities while maximizing exploration of the contract itself.
 
\section{Discussion}
\label{sec:discussion}
In this section, we discuss the limitations and potential improvements of \ToolName{}.

\PP{LLM Generation Analysis.} The generation of \ToolName{} sequences depends on LLMs, introducing three key considerations:

\textit{Contract Scale Limitations.} \quad
For <1KB contracts ($\sim${}2 functions), all fuzzers can achieve high coverage (90\%+), as optimal sequences are simple and shallow (excluding random strategy). Large contract sizes may exceed LLM context limits, harming our framework's usability. We conduct quantitative experiments in three datasets to examine how the contract size affects LLM response success rates. As shown in Table~\ref{tab:token_size}, success rates remain high (92\%+) for contracts up to 50KB (12,000 tokens), but drop significantly to 75\% for contracts exceeding 70KB (22,000 tokens) due to context limits. This indicates the need for contract segmentation strategies, hierarchical analysis approaches, or integration with more capable LLM architectures to handle enterprise-scale contracts.

\begin{table}[b]
\vspace{-11pt}
\centering
\caption{\small Contract size with its mean number of instructions impacts the success rate of LLM response}
\label{tab:token_size}
\vspace{-10pt}
{\fontsize{7.8}{10.5}\selectfont
\setlength{\tabcolsep}{3pt}%
\renewcommand{\arraystretch}{1}%
\begin{tabular}{|c|c|c|c|c|c|c|}
\hline
\textbf{Contract Size} & \textless{}6KB & $\sim$10KB  & 10-30KB     & 30-50KB      & 50-70KB      & \textgreater{}70KB   \\ \hline
\textbf{Token Size}    & $\sim$1,150    & $\sim$2,450 & $\sim$4,200 & $\sim$9,000  & $\sim$12,000 & \textgreater{}22,000 \\ \hline
\textbf{Instructions}  & $\sim$3,000    & $\sim$4,500 & $\sim$7,000 & $\sim$15,000 & $\sim$22,000 & \textgreater{}45,000 \\ \hline
\textbf{Success Rate}  & 99\%           & 99\%        & 94\%        & 92\%         & 83\%         & 75\%                 \\ \hline
\end{tabular}%
}
\vspace{-3pt}
\end{table}

\textit{LLM Overhead Analysis.} \quad
LLM-based sequence generation introduces computational overhead that impacts overall fuzzing efficiency. We measured execution times across our experimental datasets comparing \ToolName{} against baseline approaches as shown in Table~\ref{tab:overhead}. While \ToolName{} requires 838s compared to MuFuzz's 603s, the LLM overhead accounts for only 28\% of total execution time, yet achieves 1.6$\times$ improvement in vulnerability detection (376 vs. 231). Future optimizations could explore lightweight LLM architectures, caching mechanisms, or parallel processing strategies to reduce latency while maintaining generation quality.

\begin{table}[t]
\vspace{3pt}
\centering
\caption{\small Average time overhead for fuzzers to test one contract. IR-Fuzz, MuFuzz and EchoFuzz all have pre-fuzz and main-fuzz stages.}
\label{tab:overhead}
\vspace{-7pt}
{\fontsize{7.4}{8.5}\selectfont
\setlength{\tabcolsep}{3pt}%
\renewcommand{\arraystretch}{1}%
\begin{tabular}{llc}
\toprule
\multicolumn{1}{l}{\textbf{Fuzzer}} & \multicolumn{1}{c}{\textbf{Overhead of Each Step (s)}} & \textbf{Total (s)} \\ 
\midrule
sFuzz                               & Main: 300                                              & 300                \\
IR-Fuzz                             & Setup: 2 + Pre: 300 + Main: 300                        & 602                \\
MuFuzz                              & Setup: 3 + Pre: 300 + Main: 300                        & 603                \\
\ToolName                           & Chain-guided: 76 + Pre: 300 + Iteration: 162 + Main: 300 & 838                \\
\bottomrule
\end{tabular}%
}
\vspace{-10pt}
\end{table}

\textit{LLM Effectiveness Conditions.} \quad
LLM advantages over static analysis emerge when the model's performance sufficiently handles contract complexity. As demonstrated in Figure~\ref{fig:VFCS-diffLLM}, in medium-scale contracts, four out of five LLMs notably outperform static analysis baselines, while in large-scale contracts, only two LLMs demonstrate superior performance. When LLM capabilities are adequate, they provide distinct clear advantages over pure static analysis methods. Future work could focus on pre-training or fine-tuning more specialized LLMs to enhance their proficiency specifically for smart contract domains.

\PP{Data Dependency Analysis} 
\ToolName{} uses LLMs to generate sequences, different from traditional approaches. 
However, current data dependency analysis relies on traditional methods, potentially missing nuanced LLM-generated sequence dependencies.
Additionally, function parameter selection corresponds to seed in the fuzzing process, as these parameters serve as mutational targets. Effective parameter combinations that trigger new paths or vulnerabilities are retained as interesting seeds for subsequent iterations, making parameter selection decisive in vulnerability discovery.
Sequence switching timing affects parameter mutation adequacy, where appropriate timing ensures sufficient parameter space exploration. 
Future work could integrate LLMs and additional properties to guide sequence switching timing.

\PP{Sequence Diversity Analysis} \ToolName{}'s current reliance solely on fuzzing time and new path discovery to determine sequence switching, that not only constrains sequence diversity but also leads to incomplete vulnerability exploration, resulting in false negatives. 
To address this, we explore Linear Time Temporal Logic (LTL) properties to enhance sequence diversity with state transitions and access control constraints, where state transitions enforce diverse function injection under temporal conditions, and access control extends switching thresholds based on exploration sufficiency.
Experimental validation on False Negative cases from Section 5.2 shows that \ToolName{}+LTL resolves 2/2 false negative vulnerabilities with a 44\% increase in unique sequence generation (from 9 to 13), as demonstrated in Table~\ref{tab:ltl}. Currently, LTL properties are manually generated based on contract analysis, and future work could automate the extraction of temporal constraints from contract code.

\begin{table}[h]
\vspace{-7pt}
\centering
\caption{\small Performance Improvement with LTL Properties}
\label{tab:ltl}
\vspace{-10pt}
{\fontsize{7}{9.3}\selectfont
\setlength{\tabcolsep}{4pt}
\renewcommand{\arraystretch}{1}%
\begin{tabular}{|c|c|c|}
\hline
\textbf{Method} & \textbf{Number of Unique Sequences} & \textbf{False Negatives Resolved} \\ \hline
\ToolName{}     & 9                                  & 0/2                              \\ \hline
\ToolName{}+LTL & \textbf{13}                        & \textbf{2/2}                     \\ \hline
\end{tabular}%
}
\vspace{-7pt}
\end{table}

\PP{VFCS and its Property} There exists a gap between our generated candidate VFCS and the ideal VFCS properties. Our chain-guided procedure represents one specific implementation approach to approximate these ideal properties as closely as possible. Current methods focus on capturing vulnerability-triggering logical information, while determining ground truth for other properties remains extremely challenging. For example, verifying minimality requires knowing the exact minimal sequence, which is impractical without exhaustive exploration, though compromise approaches exist such as testing all subsequences of bug-triggering sequences to identify practically minimal candidates for experimental purposes.

\section{Related Work}

\label{sec:related-work}

\PP{Smart Contract Fuzzing.} 
Early tools like ContractFuzzer~\cite{jiang2018contractfuzzer} generate random inputs and monitor contract execution for signs of vulnerabilities. Although effective at a basic level, random fuzzing lacks guidance, often exploring irrelevant or unreachable paths and resulting in inefficient coverage.
Tools such as sFuzz~\cite{nguyen2020sfuzz} and Harvey~\cite{wustholz2020harvey} improve upon random fuzzing by incorporating coverage feedback to guide the generation of inputs. By focusing on unexplored paths, these fuzzers enhance efficiency. However, they often struggle to penetrate deeper contract states and may plateau after initial exploration phases.
To overcome static and dynamic approach limitations, hybrid fuzzers like ConFuzzius~\cite {torres2021confuzzius} and ILF~\cite {he2019learning} integrate symbolic execution with fuzzing to expand path exploration and generate more relevant testcases. 
For example, MuFuzz~\cite {qian2024mufuzz} uses sequence-aware and mask-guided seed mutation to reach deeply nested states overlooked by others
\ToolName{} focuses on the target of ideal VFCS, leveraging LLMs for contract-specific logical extraction to boost precision and cut exploration costs, while iterating on real-time fuzzing data to purposefully guide untapped path exploration.

\PP{LLM for Fuzzing}
The integration of LLMs into fuzzing has demonstrated significant potential across domains. For example, LLM4Fuzz~\cite{shou2024llm4fuzz} leverages LLMs to prioritize high-risk code regions in smart contracts, achieving notable efficiency gains in vulnerability detection through guided exploration. Similarly, TitanFuzz~\cite{deng2023large} employs generative and infilling LLMs to synthesize input programs for deep learning libraries, surpassing traditional fuzzers in code coverage by up to 50\%. These works highlight the capability of LLMs to reduce exploration overhead and enhance input diversity. \ToolName{} differs from existing approaches in two key aspects:

\textit{Analysis Emphasis.} \quad
LLM4Fuzz prioritizes risky code via static patterns, while \ToolName {} focuses on function call sequences by understanding function interactions and state effects.
For instance, while LLM4Fuzz relies on static analysis and LLM-generated metrics, it struggles to capture dynamic state dependencies (e.g., how \code{invest} and \code{refund} offset each other's effects in a \code{fundraiser} contract, in Section~\ref{sec:motivating-example}). \ToolName{} generates VFCS using a chain-guided procedure that analyzes state-function relationships to systematically construct vulnerability-triggering sequences.

\textit{Bottleneck Handling.} \quad
While LLM4Fuzz enhances coverage and vulnerability detection via the initial input generated by LLM, it faces challenges to further improve coverage and detect additional vulnerabilities after the fuzzing process reaches a bottleneck. \ToolName{} introduces an LLM-guided refinement loop that iteratively optimizes sequence based on execution feedback. This enables dynamic adaptation to complex contract behaviors, crucial for overcoming exploration bottlenecks in stateful environments.

\section{Conclusion}

\label{sec:conclusion}

In this work, we present \ToolName, a novel LLM-guided fuzzing framework designed specifically for smart contracts. By leveraging the logical reasoning capabilities of LLMs, \ToolName{} effectively identifies candidate VFCS -- a critical component for efficient smart contract fuzzing. 
The framework first employs a chain-guided procedure that mimics the reasoning of expert auditors. 
This approach combines static analysis with few-shot prompting to enable the LLM to interpret contract behavior and propose high-quality initial VFCS candidates.
Additionally, it employs a feedback-driven iterative process where the LLM analyzes real-time fuzzing results to adaptively refine and generate new VFCS candidates.
This dynamic guidance directs the fuzzer to unexplored, vulnerable branches.
Experiments demonstrate that \ToolName{} significantly outperforms existing approaches, achieving notable improvements in both coverage and vulnerability detection.

\begin{acks}
We would like to thank Jie Liang, Yu Jiang, Luke Ong, and Andrzej Murawski for their valuable feedback and support. We are also grateful to the anonymous reviewers for their constructive comments.
This work was supported in part by the National Key Research and Development Project under Grant No. 2024YFF1401300, the Fundamental Research Funds for the Central Universities KG16358401, NSFC Program No. 62302256, No. 62502254 and NRF RSS Scheme NRF-RSS2022-009.
\end{acks}

\bibliographystyle{ACM-Reference-Format}
\bibliography{reference}

\end{sloppypar}
\end{document}